\documentclass[noline, journal abbreviation, manuscript]{copernicus}
\makeatletter
\nolinenumbers
\renewcommand\linenumbers{}
\makeatother

\begin{document}

\title{The Oxygen Valve on Hydrogen Escape Since the Great Oxidation Event}

% \Author[affil]{given_name}{surname}

\author[1]{Gregory J. Cooke}
\author[2]{Dan R. Marsh}
\author[2]{Catherine Walsh}
\author[2]{Felix Sainsbury-Martinez}
\author[3]{Marrick Braam}

\affil[1]{Institute of Astronomy, University of Cambridge, CB3 0HA, UK.}

\affil[2]{School of Physics and Astronomy, University of Leeds, Leeds, LS2 9JT, UK.}

\affil[3]{Center for Space and Habitability, University of Bern, Gesellschaftsstrasse 6, 3012 Bern, Switzerland.}

%% The [] brackets identify the author with the corresponding affiliation. 1, 2, 3, etc. should be inserted.

%% If an author is deceased, please mark the respective author name(s) with a dagger, e.g. "\Author[2,$\dag$]{Anton}{Smith}", and add a further "\affil[$\dag$]{deceased, 1 July 2019}".

%% If authors contributed equally, please mark the respective author names with an asterisk, e.g. "\Author[2,*]{Anton}{Smith}" and "\Author[3,*]{Bradley}{Miller}" and add a further affiliation: "\affil[*]{These authors contributed equally to this work.}".

\correspondence{Gregory Cooke (gjc53@cam.ac.uk), Catherine Walsh (c.walsh1@leeds.ac.uk)}

\runningtitle{The Oxygen Valve on Hydrogen Escape Since the Great Oxidation Event}

\runningauthor{Cooke}

\received{}
\pubdiscuss{} %% only important for two-stage journals
\revised{}
\accepted{}
\published{}

%% These dates will be inserted by Copernicus Publications during the typesetting process.

\firstpage{1}

\maketitle

\begin{abstract}
The Great Oxidation Event (GOE) was a $200$ Myr transition circa 2.4 billion years ago that converted the Earth's anoxic atmosphere to one where molecular oxygen (O$_2$) was abundant (volume mixing ratio $>10^{-4}$). This significant rise in O$_2$ is thought to have substantially throttled hydrogen (H) escape and the associated water (H$_2$O) loss. Atmospheric estimations from the GOE onward place O$_2$ concentrations ranging between 0.1\% to 150\% PAL, where PAL is the present atmospheric level of 21\% by volume. In this study we use WACCM6, a three-dimensional Earth System Model to simulate Earth's atmosphere and predict the diffusion-limited escape rate of hydrogen due to varying O$_2$ post-GOE. We find that O$_2$ indirectly acts as a control valve on the amount of hydrogen atoms reaching the homopause in the simulations: less O$_2$ leads to decreased O$_3$ densities that reduce local tropical tropopause temperatures by up to 17 K, which increases H$_2$O freeze-drying and thus reduces the primary source of hydrogen in the considered scenarios. The maximum differences between all simulations in the total H mixing ratio at the homopause and the associated diffusion-limited escape rates are a factor of 3.2 and 4.7, respectively. The prescribed CH$_4$ mixing ratio (0.8 ppmv) sets a minimum diffusion escape rate of $\approx 2 \times 10^{10}$ mol H yr\textsuperscript{-1}, effectively a negligible rate when compared to pre-GOE estimates ($\sim10^{12}-10^{13}$ mol H yr\textsuperscript{-1}). Because the changes in our predicted escape rates are comparatively minor, our numerical predictions support geological evidence that the majority of Earth's hydrogen escape occurred prior to the GOE. Our work demonstrates that estimations of how the tropical tropopause layer and the associated hydrogen escape rate evolved through Earth's history requires 3D chemistry-climate models which include a global treatment of water vapour microphysics.
\end{abstract}

\introduction[Introduction]  %% \introduction[modified heading if necessary]
\label{Introduction section}

A key component of Earth's atmospheric and biological history is its water (H$_2$O). The relative amount of H$_2$O that was present during the Earth's formation versus the quantity delivered by volatile-rich bodies \citep[e.g., comets and asteroids;][]{2000Icar..148..508D, 2013ApJ...767...54I}, as well as the H$_2$O lost to space through hydrogen escape \citep{2001Sci...293..839C, 2008Icar..194...42G}, or exchanged into the interior \citep{2015Sci...350..795H, 2017RSPTA.37550393K}, has had an important impact on how the Earth's surface and the atmosphere have developed. Beyond Earth, the solar system's terrestrial planet atmospheres are desiccated and consequently, Venus and Mars are both apparently desolate environments. It's possible that Venus was never habitable \citep{2021Natur.598..276T, 2023PNAS..12009751W, 2024NatAs.tmp..289C}, but Mars was likely habitable early in its history before its water was isolated to the subsurface and the crust \citep{2010NatGe...3..459D, 2018Sci...361..490O, 2021Sci...372...56S, 2021NatAs...5...63L, 2022NatAs...6.1263S, 2024RemS...16..824Z, wright2024liquid}. Venus is now too hot for liquid H$_2$O to persist and probably lost much of its natal H$_2$O through atmospheric hydrogen escape \citep{1969JAtS...26.1191I, 2006P&SS...54.1425K, 2019JGRE..124.2015K, 2024AGUFMSM13E..08C, Gu_Venus_Photo_2025}.

In contrast, Earth has mostly remained temperate throughout its history, with liquid water persisting on the surface for over 4 billion years \citep[and references therein]{2012RvGeo..50.2006F}, although there were periods of widespread glaciation at either end of the Proterozoic eon \citep[2.4 -- 0.541 Gyr ago;][]{young2013precambrian, 2020PNAS..11713314W}. A mostly temperate Earth has been possible despite a fainter Sun in Earth's past \citep{bahcall2001solar, 2012RvGeo..50.2006F, 2020SSRv..216...90C}, in combination with fluctuations in albedo \citep{2020SSRv..216...90C, goldblatt2021earth} and atmospheric composition, in particular the greenhouse gases CO$_2$ \citep{2006PreR..147..148S, 2013ChGeo.362..224S, kanzaki2015estimates} and CH$_4$ \citep{daines2016effect, zhao2018terrestrial, 2019E&PSL.522...48L, fakhraee2019proterozoic}. Geochemical proxies suggest a broad range but generally greater past abundance for CO$_2$, whereas only loose constraints have been placed on CH$_4$ \citep[and references therein]{2020SciA....6.1420C}, with debates regarding whether CH$_4$ during the Proterozoic was higher or lower in abundance than present day still ongoing (see Discussion section).

As all known life requires liquid H$_2$O, the history of Earth's H$_2$O is inextricably tied to the evolution of Earth's life. From isotopic evidence and computational modelling, since its formation, the Earth may have lost between 0.13 and 2 times the present ocean volume \citep{2012PNAS..109.4371P, 2018E&PSL.497..149K, 2019GeCoA.244...56Z}. Indeed, the early Archean Earth (4 -- 2.4 Gyr ago), before 3.2 Gyr ago, may have been fully covered in a global ocean \citep{2021AGUA....200323D}. The current hydrogen escape rate is $\approx 4\times10^{10}$ mol of H $\textrm{yr}^{-1}$, corresponding to roughly 1 metre of global water loss per 1 billion years \citep{2013ChGeo.362...26Z}. Given that the Earth's ocean can be up to 11 km deep \citep{gardner2014so}, and its mean depth is $\approx 3.6$ km \citep{charette2010volume}, the loss rate is insignificant. This suggests a different past for hydrogen escape than the constant modern rate would imply.

Therefore, an important but uncertain component of Earth's water loss is determining the amount of hydrogen (H) that has escaped to space. As the lightest element, H escapes atmospheres more readily than other elements and chemical compounds. On Earth, hydrogen atoms exist in many different chemical species in the lower and middle atmosphere \citep{2005ama..book.....B}. The four major chemical constituents that contribute hydrogen atoms to the upper atmosphere, and therefore the species we focus on in this work, are H, H$_2$, H$_2$O, and CH$_4$. Atmospheric escape of hydrogen can occur from a planet through several mechanisms, including: Jeans escape, hydrodynamic flow, sputtering, impacts, photochemical escape, and charge exchange escape \citep{2007SSRv..129..245L, 2010ApJ...716.1573J, 2018PhyU...61..217S, 2020JGRA..12527639G}. For these mechanisms, with the exception of impacts and hydrodynamic flow, the hydrogen escapes from the exosphere at the top of the atmosphere. In such cases, hydrogen cannot escape faster than it can be delivered to the exosphere through molecular diffusion. This is known as `diffusion-limited hydrogen escape' \citep{1973JAtS...30.1481H, 1976AREPS...4..265H, 1977evat.book.....W, 2003ARA&A..41..429K, 2013ChGeo.362...26Z}. The diffusion-limited rate of hydrogen escape is set by two controlling bottlenecks, the upward diffusion of hydrogen atoms (mostly in H$_2$O and CH$_4$) through the tropopause `cold trap', and the upward diffusion of hydrogen atoms through molecular diffusion above the homopause\footnote{The homopause is the atmospheric point above which molecular diffusion dominates and below which turbulent diffusion dominates} \citep{2001Sci...293..839C, catling2017atmospheric}. 

The primary hydrogen bottleneck, the cold trap, is where the saturation mixing ratio of H$_2$O and ice is at a minimum, such that it modulates the amount of H$_2$O propagating upwards into the stratosphere \citep{2015ApJ...813L...3K}. On Earth, the cold trap is typically described as the tropical tropopause layer (TTL). The TTL is three-dimensional and is defined as the transition region between the Hadley circulation in the troposphere and the start of the deep branch of the Brewer-Dobson circulation \citep{2005ama..book.....B, 2009RvGeo..47.1004F, 2017ACP....17.4337G}. Despite being located above the tropics (at 12 -- 16 km in altitude, which is 70 -- 150 hPa in pressure), which have on average greater surface temperatures than higher latitudes, the TTL is defined by its cold temperatures \citep{2005ama..book.....B, 2009RvGeo..47.1004F}. The cause of this is adiabatic cooling as air is transported upwards from the convective troposphere into the stratosphere, and this cold temperature region leads to the removal of water vapor from the atmosphere through condensation \citep{1949QJRMS..75..351B, newell1981stratospheric, 1982GeoRL...9..605D, 2007ACPD....7.8933P}. The location, temperatures, and dynamics of the TTL are affected by seasons \citep{kim2012tropical} and climate change \citep{2013NatGe...6..169R, 2017JCli...30.1245L, 2023FrEaS..1177502Z, 2024ACP....24.7405Z}. Of the major H-bearing species, only H$_2$O is significantly affected by temperatures in the TTL, whereas H$_2$ and CH$_4$, which don't condense in Earth's atmosphere, are unaffected by the cold trap mechanism.

In the stratosphere, hydrogen atoms are interchanged between CH$_4$, H$_2$O and H$_2$, until reaching the homopause \citep{catling2017atmospheric}. At altitudes exceeding the homopause, through diffusive separation, the lighter atmospheric constituents with lower molecular masses increase in relative abundance with decreasing pressure (e.g., H, He, O, and N). Note that H$_2$ would also increase but it is rapidly photodissociated into H. Hence, above the homopause, H dominates the total hydrogen budget \citep{catling2017atmospheric}. 

The atmosphere of the Archean Earth was likely weakly reducing with greater quantities of CH$_4$, CO, and H$_2$ \citep{2001OLEB...31..271K, kharecha2005coupled, 2019GeCoA.262..207K, 2020SciA....6.1420C}. Midway through Earth's history, approximately 2.4 billion years ago, a rise in oxygen known as the Great Oxidation Event \citep[GOE;][]{2000Sci...289..756F, 2002GeCoA..66.3811H, 2019PNAS..11617207H, 2020PNAS..11713314W, 2021NatGe..14..143G} has been proposed to have significantly reduced hydrogen escape from $\sim10^{12}-10^{13}$ mol H yr\textsuperscript{-1} to $\sim10^{10}$ mol H yr\textsuperscript{-1} \citep{2001Sci...293..839C, 2013ChGeo.362...26Z}. This is because an oxygenated atmosphere destroys reducing species (e.g., H$_2$ and CH$_4$), resulting in lower amounts of such gases which contribute hydrogen atoms to the upper atmosphere.

In conjunction, atmospheric escape of hydrogen may have played a significant role in shaping Earth's atmosphere through time. It is thought to at least be partially responsible for oxidising the Earth's surface: when H$_2$O is photodissociated above the troposphere, the liberated hydrogen (H) can escape to space, and the oxygen (O) left behind oxidises the Earth \citep{2001Sci...293..839C, claire2006biogeochemical, 2013ChGeo.362...26Z}. Alongside photosynthetic production of O$_2$ from cyanobacteria, this process is hypothesized to have contributed to the occurrence of the GOE because the process of hydrogen escape and subsequent oxidation is irreversible, leading to a reduced sink of O$_2$ over time \citep{2001Sci...293..839C, claire2006biogeochemical, 2013ChGeo.362...26Z}. Nevertheless, considerable uncertainty persists regarding the precise temporal evolution of Earth's atmosphere \citep{kasting2025evolution}, with hydrogen escape a key aspect of the holistic picture.

There is no direct way to determine the amount of hydrogen that has escaped in Earth's history. As mentioned, several measurements can be used to infer this quantity, and the chronology of one of the heaviest gases, Xe, may be informative. Xe is more easily ionised compared to the other noble gases which do not exhibit the same fractionation \citep[the so-called Xenon paradox;][]{1977Sci...198..453A, 2013NatCh...5...61Z}. It was proposed that Xe$^+$ could escape Earth's atmosphere when propelled by H$^+$ ions, and this has been shown to be a feasible physical process \citep{2019GeCoA.244...56Z}. Accordingly, the timing of Earth's Xe fractionation \citep[see figure 3 in][]{2019GeCoA.244...56Z} may trace high levels of atmospheric hydrogen escape during the Archean eon. There has been relatively small changes in Xe fractionation since the GOE \citep{2018GeCoA.232...82A}, indicating the vast majority of historical hydrogen escape occurred prior to this time (i.e., before $\sim2.4$ Gyr ago).

Whilst hydrogen escape rates may have been much lower in the last $\sim2.4$ Gyr, to our knowledge, no work has quantitatively estimated escape rates since the GOE occurred using a 3D model. In this work, we build upon a previous study \citep{2022RSOS....911165C} which simulated a range of O$_2$ concentrations that may have been present on the Earth for the last 2.4 Gyr and found that lower O$_3$ columns were predicted compared to previous work \citep{2003ARA&A..41..429K, 2003AsBio...3..689S, 2015ApJ...806..137R, 2018ApJ...854...19R, 2020ApJ...904...10K, 2023RSOS...1030056J}. More recent work has in part supported the results of lower O$_3$ for a given O$_2$ mixing ratio \citep{2022CliPa..18.2421Y,  ji2024correlated, liu2025evolution}. O$_3$ and its associated UV heating is responsible for the temperature inversion that creates the Earth's stratosphere, and this heating affects the temperatures in the TTL. Therefore, the O$_2$ fluctuations following the GOE may have modulated the upward transport of H$_2$O through the TTL and consequently, hydrogen escape too.

Using the 3D chemistry-climate model WACCM6, we first explore the affect of changing the O$_2$ abundance on the TTL region, before investigating the influence it may have had on hydrogen escape during the Proterozoic eon. Our aim is to estimate how changes to atmospheric O$_2$ alone can affect the diffusion-limited hydrogen escape rate. In terms of hydrogen escape fluctuations, our focus is on H$_2$O, and we hold the mixing ratios of H$_2$ and CH$_4$ constant at the surface. We note that CO$_2$, CH$_4$ and H$_2$ are likely to have varied during the past 2.4 billion years \citep{2020SciA....6.1420C}, as will have the luminosity of the Sun \citep{2012RvGeo..50.2006F}. These variables are important because CH$_4$ and H$_2$ are key hydrogen carriers to the upper atmosphere, and additionally CH$_4$ and CO$_2$ provide greenhouse warming which can affect TTL temperatures. The TTL will also have been affected from the reduced total insolation from a younger Sun which would act to cool the troposphere. The purpose of this work is, however, to present the mechanism by which O$_2$ affects hydrogen escape. We calculate such H escape deviations, paving the way for future work on the topic to investigate other physical, chemical, and biological influences.

\section{Numerical methods}
\subsection{WACCM6 model}

The simulations we analyse are those from \cite{2022RSOS....911165C}, where the Earth System Model WACCM6 \citep{2019JGRD..12412380G} was employed to model different oxygenation levels. WACCM6 is a configuration of the Coupled Earth System Model (CESM), and we use version 2.1.3 (CESM2.1.3\footnote{\href{https://www.cesm.ucar.edu/models/cesm2}{https://www.cesm.ucar.edu/models/cesm2}}) with minor code alterations in the initial conditions\footnote{ExoCESM GitHub: \href{https://github.com/exo-cesm/CESM2.1.3/tree/main/O2_Earth_analogues}{https://github.com/exo-cesm/CESM2.1.3/tree/main/O2\textunderscore Earth\textunderscore analogues}}, which we describe in this section. 

The first simulation we perform commences with initial conditions which are intended to be representative of Earth's pre-industrial (PI) atmosphere circa 1850, alongside contemporary ocean and land settings, and the current orbital setup with a $23.4^\circ$ obliquity and 24-hour rotation rate. Deep time paleoclimate studies would often change the continents and we accept that this is a limitation with the WACCM6 model where continents cannot currently be modified. The circulation pattern and climate would have been different in the past due to various factors, which include the younger Sun, different greenhouse gas concentration, and the formation and break up of supercontinents such as Pangea and Columbia/Nuna \citep{Rogers_2002_Columbia, 2012RvGeo..50.2006F, 2017Geo....45..231F, 2020SciA....6.1420C}. The model's spatial resolution for the atmosphere was set at $1.875^\circ$ latitude by $2.5^\circ$ longitude (yielding a $96 \times 144$ horizontal grid). This configuration utilizes 70 atmospheric layers in vertical pressure coordinates extending from the surface (approximately 1000 hPa) to the thermosphere ($4.5 \times 10^{-6}$ hPa, or $\sim 150$ km). The land component (CLM5.0) operates on the same $1.9^\circ \times 2.5^\circ$ horizontal grid as the atmosphere to maintain consistency in surface flux exchanges.  The simulation employs an interactive ocean via the POP2 model, which operates on a $1^\circ$ displaced-pole grid. This ocean grid features $320 \times 384$ points, with enhanced latitudinal resolution near the equator to better resolve tropical dynamics. This same grid is utilized by the Sea-Ice (CICE5) and Wave (WaveWatch3) components. Additionally, freshwater discharge is included in the MOSART river transport model on a $0.5^\circ$ global grid. 

Both atmospheric and oceanic components of the model are interactive, allowing dynamic responses to environmental stimuli like temperature changes. Our simulations incorporated middle atmosphere chemistry as described by \cite{2020JAMES..1201882E}, featuring 98 chemical species and 298 reactions, including O$_3$ chemistry. The atmospheric time step, $\Delta t$, was set at 30 minutes. The concentrations of 75 species were computed using an implicit numerical scheme, while 22 long-lived species were computed using an explicit numerical scheme \citep{sandu1997benchmarking,2005ama..book.....B}. N$_2$ is considered invariant. Absorption of light by CO$_2$ and H$_2$O in the Schumann-Runge (S-R bands) is not included in these simulations, and neither is scattering at wavelengths $<200$ nm. For more details on the inclusion of these processes, we refer the reader to \cite{ji2023comparison} and \cite{ji2024correlated}, as well as Section~\ref{Model limitations section} where we discuss how it might impact our results.

In total, eight simulations were performed for the Earth system with different mixing ratios of atmospheric O$_2$, where the atmospheric pressure was kept the same by changing the abundance of the background gas, N$_2$. O$_2$ mixing ratios are given in terms of present atmospheric level (PAL), which is 21\% by volume. The O$_2$ surface mixing ratios span values between 0.1 -- 150\% PAL, which are within estimates for the Earth over the last 2.4 billion years \citep{2014Natur.506..307L, 2019MinDe..54..485L, 2020SciA....6.1420C, 2020PreR..34305722S}. These temporal estimates do not always agree, so we remain agnostic and perform simulations covering the whole range. The simulation names are PI, 150\% PAL, 50\% PAL, 10\% PAL, 5\% PAL, 1\% PAL, 0.5\% PAL, and 0.1\% PAL. CO$_2$, CH$_4$, N$_2$O, and H$_2$, were held fixed at values of 284 ppmv, 0.8 ppmv, 273 ppbv, and 0.5 ppmv, respectively. The mixing ratios or fluxes of gases specified at the surface were kept constant as O$_2$ was altered. This assumption is likely not realistic for the entirety of the last 2.4 Gyr, but it allows us to isolate the role O$_2$ changes make to the chemistry and dynamics of the atmosphere. Water vapor feedback, in terms of evaporation and rainout, is included in WACCM6, but hydrogen escape and water loss are not explicitly simulated. Instead, we estimate these values once quasi-steady state has been reached in the simulation (see section \ref{Diffusion-limited hydrogen escape rate section}).

We average the outputs of the WACCM6 simulations over the last four years of the simulation once the middle atmospheric \citep[defined here as between 12 -- 100 km in altitude;][]{2005ama..book.....B} trend in total hydrogen atoms has halted, and this required between 30 and 90 simulations years for the different scenarios.

\subsection{Diffusion-limited hydrogen escape rate}
\label{Diffusion-limited hydrogen escape rate section}

The diffusion-limited hydrogen escape rate \citep{1973JAtS...30.1481H, 2003ARA&A..41..429K}, $\Phi_\textrm{esc}$, is proportional to the total mixing ratio of hydrogen-bearing species, $f_\textrm{T}(\textrm{H})$, at the homopause: 

\begin{equation}
    \Phi_\textrm{esc} \propto f_\textrm{T}(\textrm{H}),
    \label{Equation 1}
\end{equation}

\noindent where $f_\textrm{T}(\textrm{H})$,  can be written as

\begin{equation}
    f_\textrm{T}(\textrm{H}) = f(\textrm{H}) + 2f(\textrm{H}_2) + 2f(\textrm{H}_2\textrm{O}) + 4f(\textrm{CH}_4) ...,
    \label{Equation 2}
\end{equation}

\noindent and so on, accounting for all hydrogen-bearing species. We focus on evaluating the abundances of H, H$_2$, H$_2$O, and CH$_4$. In modern Earth's atmosphere, at the tropopause, the H$_2$ and CH$_4$ mixing ratios are $\approx0.55$ ppmv \citep{catling2017atmospheric} and $\approx1.9$ ppmv \citep{Saunois_Methane_2025}, respectively. There, the H$_2$O mixing ratio varies between $\approx 2.5-4.5$ ppmv, with a mean stratospheric entry value of $\sim3.5-4$ ppmv \citep{2009RvGeo..47.1004F, 2019JGRD..12412380G}. This puts the stratospheric entry value of total hydrogen at $\sim 17$ ppmv, although with CH$_4$ at a lower concentration during the pre-industrial period, the entry value would be closer to $\sim 12$ ppmv. In our simulations, the lower boundary mixing ratios are instead mixing ratios of 0.5 ppmv and $0.8$ ppmv for H$_2$ and CH$_4$, respectively, corresponding to pre-industrial values. Consequently, H$_2$O is the primary carrier of hydrogen atoms to the middle atmosphere. Other H-bearing species such as OH, HO$_2$, and H$_2$O$_2$ are included in the WACCM6 model but are not sufficiently abundant to contribute significantly to the overall hydrogen budget.

The calculation of $\Phi_\textrm{esc}$ at the homopause can be made by considering the binary diffusion coefficients which depend on the temperature of the atmosphere and the chemical species. $\Phi_\textrm{esc}$, in units of atoms cm\textsuperscript{-2} s\textsuperscript{-1}, is given by 

\begin{equation}
    \Phi_\textrm{esc} = \sum_{i=1}^{n}  \frac{N_i f(i) b_i}{H_h}, \label{Equation 3}  
\end{equation}

\noindent where $H_h$ is the atmospheric scale height at the homopause, $N_i$ is the number of hydrogen atoms in each chemical species, $i$, and $b_{\textrm{i}}$ is the binary diffusion parameter for each constituent. The binary diffusion coefficient depends on the temperature of the atmosphere at the homopause ($T_h$), the masses of the species, and their collision cross-sections. For each species considered here, the binary diffusion coefficients are $b_{\textrm{H}} = 6.5 \times 10^{17} T_{h}^{0.7}$, $b_{\textrm{H}_2} = 2.67 \times 10^{17} T_{h}^{0.75}$, $b_{\textrm{H}_2\textrm{O}} = 0.137 \times 10^{17} T_{h}^{1.072}$, $b_{\textrm{CH}_4} = 0.734 \times 10^{17} T_{h}^{0.750}$. These binary diffusion parameters for H, H$_2$, and H$_2$O are sourced from \cite{1973JAtS...30.1481H} and CH$_4$ is sourced from \cite{banks1973aeronomy}. The calculations are made for the whole atmosphere (latitude and longitude at the homopause) before calculating a longitudinally-averaged and latitudinally-weighted mean.

From Earth's exosphere, hydrogen escapes faster than it is delivered, such that throughout this work, we will assume that the loss rate of hydrogen atoms is diffusion-limited (ignoring hydrodynamic escape and impacts). Note that it is possible for hydrogen to escape slower than this upper limit, provided the rate of loss from the exosphere is the limiting factor \citep[e.g., energy-limited escape, photon-limited escape;][]{2005Sci...308.1014T,2016A&A...585L...2S, 2016ApJ...816...34O, 2017ApJ...845..130L}.

\section{Results}

% Zonal mean temperature figure
\begin{figure*}[b!]
	\centering
	\includegraphics[width=1\textwidth]{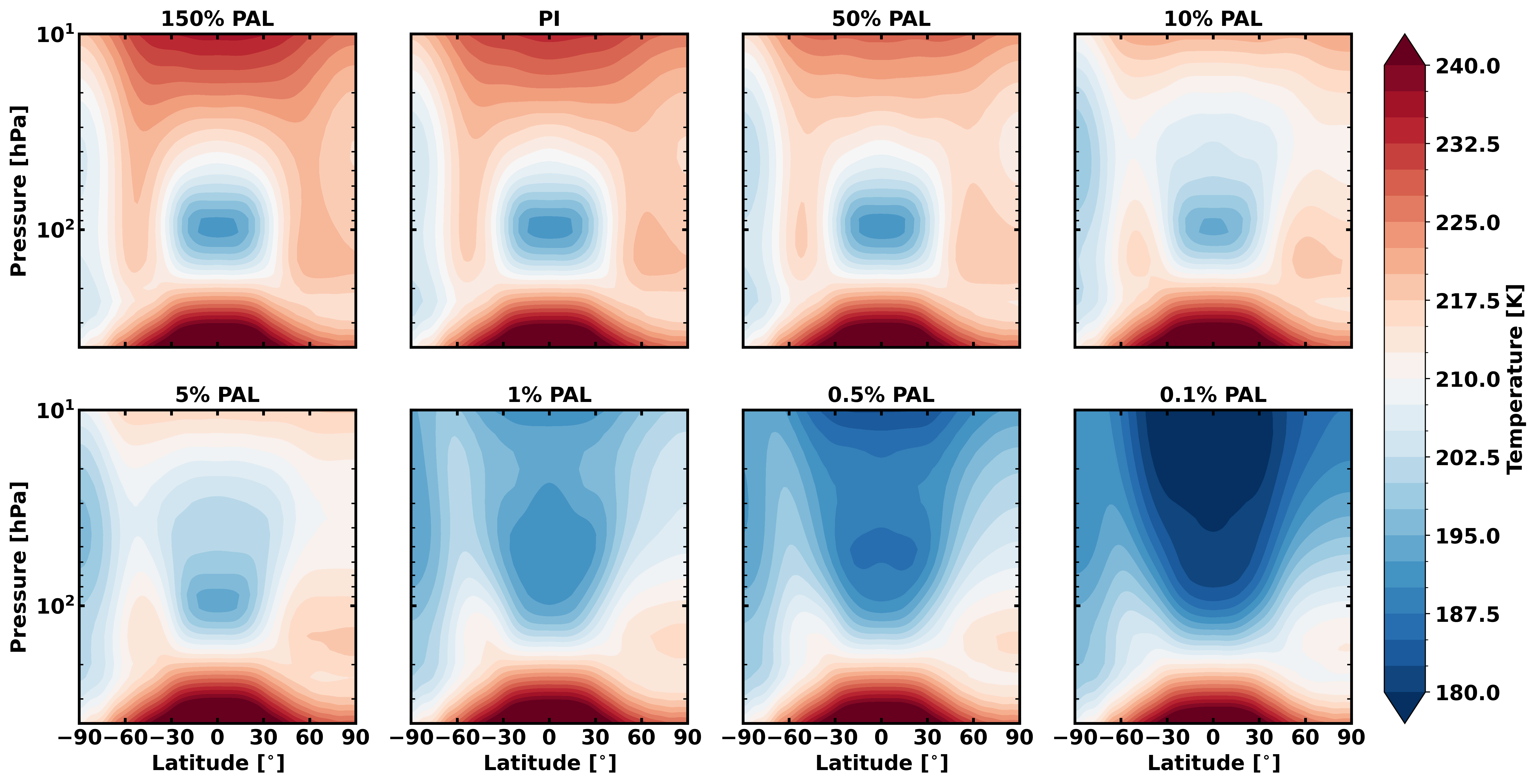}
    \caption{The zonal mean of atmospheric temperature (in K) is shown for simulations with O$_2$ mixing ratios between 150\% PAL and 0.1\% PAL, where O$_2$ decreases in abundance from the top left panel to the bottom right panel. The colour bar shows hotter temperatures in red and colder temperatures in blue. The TTL region is $\pm23.4^\circ$ from the equator and between 70–150 hPa in pressure.} 
    \label{Zonal mean temperature figure}
\end{figure*}

\label{Results section}

As part of the hydrological cycle, liquid water evaporates from the surface and warm, moist air rises due to buoyancy. Hence, rivers, lakes, and predominantly the ocean, act as a sources of hydrogen atoms to the atmosphere. We will present the results in terms of how the total hydrogen mixing ratio starts at the surface and how it changes with decreasing atmospheric pressure (increasing altitude). This includes tracking the proportion of hydrogen atoms bound in different species, such as H$_2$O and CH$_4$. We will show how the O$_2$ mixing ratio affects the total hydrogen mixing ratio through the atmosphere and ultimately the amount of hydrogen atoms diffusing up to the exosphere. In the following subsection, we consider how O$_2$ affects clouds, condensation, and the cold trap at the top of the troposphere. These processes act to keep hydrogen atoms in the troposphere that would otherwise mix into the stratosphere and upwards to eventually escape to space.

\subsection{Water vapour through the cold trap}

% Oxygen versus ozone curve figure
\begin{figure}[b!]
	\centering
	\includegraphics[width=1\textwidth]{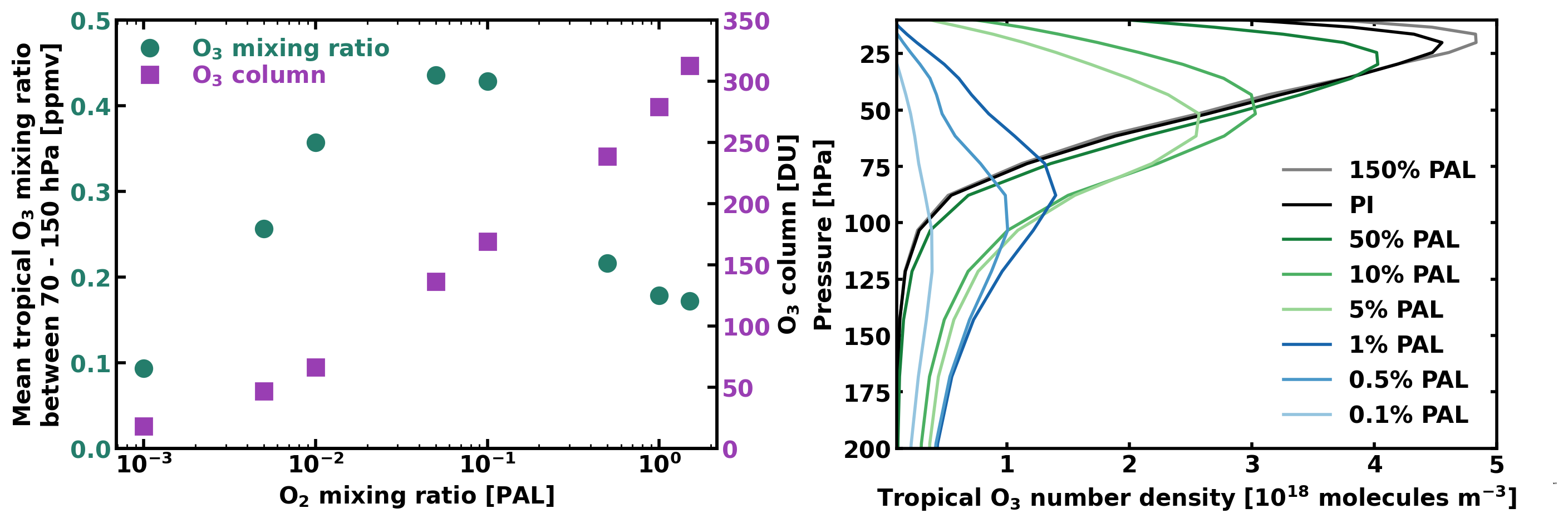}
    \caption{\textbf{Left:} The mean tropical (defined at latitudes $\pm24^\circ$ from the equator) O$_3$ mixing ratio between 70 -- 150 hPa is plotted on the left vertical axis in teal circles against the atmospheric mixing ratio of O$_2$ at the surface in terms in terms of the present atmospheric level (PAL), which is 21\% by volume. The total global ozone (O$_3$) column density in Dobson Units (DU, where 1 DU $= 2.69\times10^{20}$ molecules m\textsuperscript{-2}) is also plotted in purple squares on the right vertical axis against the atmospheric mixing ratio of O$_2$. \textbf{Right}: The O$_3$ number density is shown for all eight simulations between 10 -- 200 hPa on a linear scale \cite[note the number density does not go to zero at pressures greater than 200 hPa, see][]{2022RSOS....911165C}.}
    \label{Oxygen versus ozone curve figure}
\end{figure}

In every simulation, the global mean mixing ratio of each species at the surface for H$_2$O, H, CH$_4$, and H$_2$, is $\sim 10,000$ ppmv, $<10^{-8}$ ppmv, $\approx0.8$ ppmv, and $\approx0.5$ ppmv respectively, with the latter two molecules specified with fixed mixing ratios at the surface. As temperature and pressure decline with altitude in the troposphere, the atmosphere cannot hold as much H$_2$O vapor. This is because the drop in pressure causes the air to cool (adiabatic cooling), and the saturation vapor pressure of water, which determines the air's capacity for water vapor, is strongly dependent on temperature. Fig.~\ref{Zonal mean temperature figure} presents the zonal mean (averaged over longitude) of the atmospheric temperature between the surface and 10 hPa ($\sim 30$ km in altitude) for all the WACCM6 simulations. The temperature structure of the atmosphere is set by the strength and spectral shape of the incoming radiation, the moist adiabat in the troposphere, atmospheric composition (providing heating and cooling in different layers), clouds and ice, and atmospheric dynamics. Between 150\% PAL and 50\% PAL, the temperature structure at pressures greater than 10 hPa (or altitudes below 10 hPa) is similar. At 10\% and 5\% PAL, the temperature increases in the middle atmosphere, and this is noticeable in the tropics between 200 -- 30 hPa and at northern ($45-90^\circ$) latitudes between 200 -- 100 hPa. At $\leq1$\% PAL of O$_2$, the stratosphere begins to become significantly cooler, with the 10 hPa pressure level $\sim 60$ K cooler in the 0.1\% PAL simulation compared to the 150\% PAL simulation. A globally averaged stratospheric temperature inversion of $\approx 2$ K and $\approx 1$ K exists in the 1\% PAL and 0.1\% PAL simulations, respectively, which is significantly smaller than the $\approx 60$ K temperature inversion in the PI atmosphere. 

Fig.~\ref{Oxygen versus ozone curve figure} shows the mixing ratio of atmospheric O$_2$ plotted against the O$_3$ column density, as well as the mean O$_3$ mixing ratio between 70 and 150 hPa in the tropical region, accounting for the expanse of the TTL. The figure also shows the O$_3$ number density in the tropics between 10 and 200 hPa. Whilst the global O$_3$ column density increases with increasing O$_2$ mixing ratio, this is not the case for the O$_3$ mixing ratio around the TTL region: there is instead a peak at $\approx5-10$\% PAL. The O$_3$ number density also mirrors this relationship. When O$_2$ is decreased, O$_3$ accumulates lower down in the atmosphere because the wavelengths that photolyse O$_2$ to produce ``odd oxygen'' (O + O$_3$) are able to penetrate deeper. Consequently, for O$_2$ concentrations of $\leq 10$\% PAL, O$_3$ maximizes in this region near the TTL. The TTL temperatures are a balance between shortwave heating and the emission and absorption of longwave radiation. Whilst this causes higher temperatures than the PI atmosphere for 5 -- 10\% PAL, simulations with $\leq1$\% PAL have resultant lower temperatures. This effect is a local TTL phenomenon; globally, the mean tropopause temperatures decrease with decreasing O$_2$.

These results are in line with previous O$_3$ depletion experiments (albeit of a much smaller magnitude) which found that reduced O$_3$ concentrations cooled the TTL and led to less stratospheric water vapour \citep{Forster_2007_O3,Xie_2008_O3_tape, Lu_2021_O3, Zhou_2022_Ozone}. The cooling was attributed to radiative processes. Indeed, \cite{Forster_2007_O3} showed that less O$_3$ in the tropical lower stratosphere caused cooling that extended into the upper tropical troposphere due to reduced downward longwave radiation. Similarly, here we find that lower middle stratospheric densities of O$_3$ reduces the propagation of longwave radiation downwards. This mechanism is the cause of the hotter temperatures in the higher oxygenated scenarios in Fig.~\ref{Zonal mean temperature figure} compared to scenarios with $\leq1$\% PAL. This is also why there is an asymmetry in temperatures because of hemispherical asymmetries in the O$_3$ distribution (there is more O$_3$ in the northern hemisphere) as a result of the Brewer-Dobson circulation \citep{1926RSPSA.110..660D, 1949QJRMS..75..351B, 2014RvGeo..52..157B}.

In Section \ref{Introduction section}, we described how H$_2$O moves through the TTL to get into the stratosphere on modern Earth. The cold trap mechanism is indirectly sensitive to O$_2$ concentrations resulting from the aforementioned difference in O$_3$ and hence UV heating. Due to the difference in time-averaged zonal mean TTL temperatures between the PI and 0.1\% PAL simulations of up to $\approx17$ K, more H$_2$O is frozen out in the form of water clouds, ice, and ice clouds \citep[e.g., cirrus clouds;][]{2012JGRD..117.4211W} before crossing $\approx100$ hPa in the 0.1\% PAL case compared to the standard PI case. Stratospheric O$_3$ and H$_2$O are greenhouse gases, and with less H$_2$O reaching the stratosphere as a consequence of reduced O$_3$, this decreases tropospheric heating \citep{2023CliPa..19.1201D}, causing a positive feedback effect, albeit a small one: there is some cooling at the surface, but this is limited to $< 3$ K in terms of global averaged surface temperatures.

\begin{figure}[t!]
	\centering
	\includegraphics[width=1\textwidth]{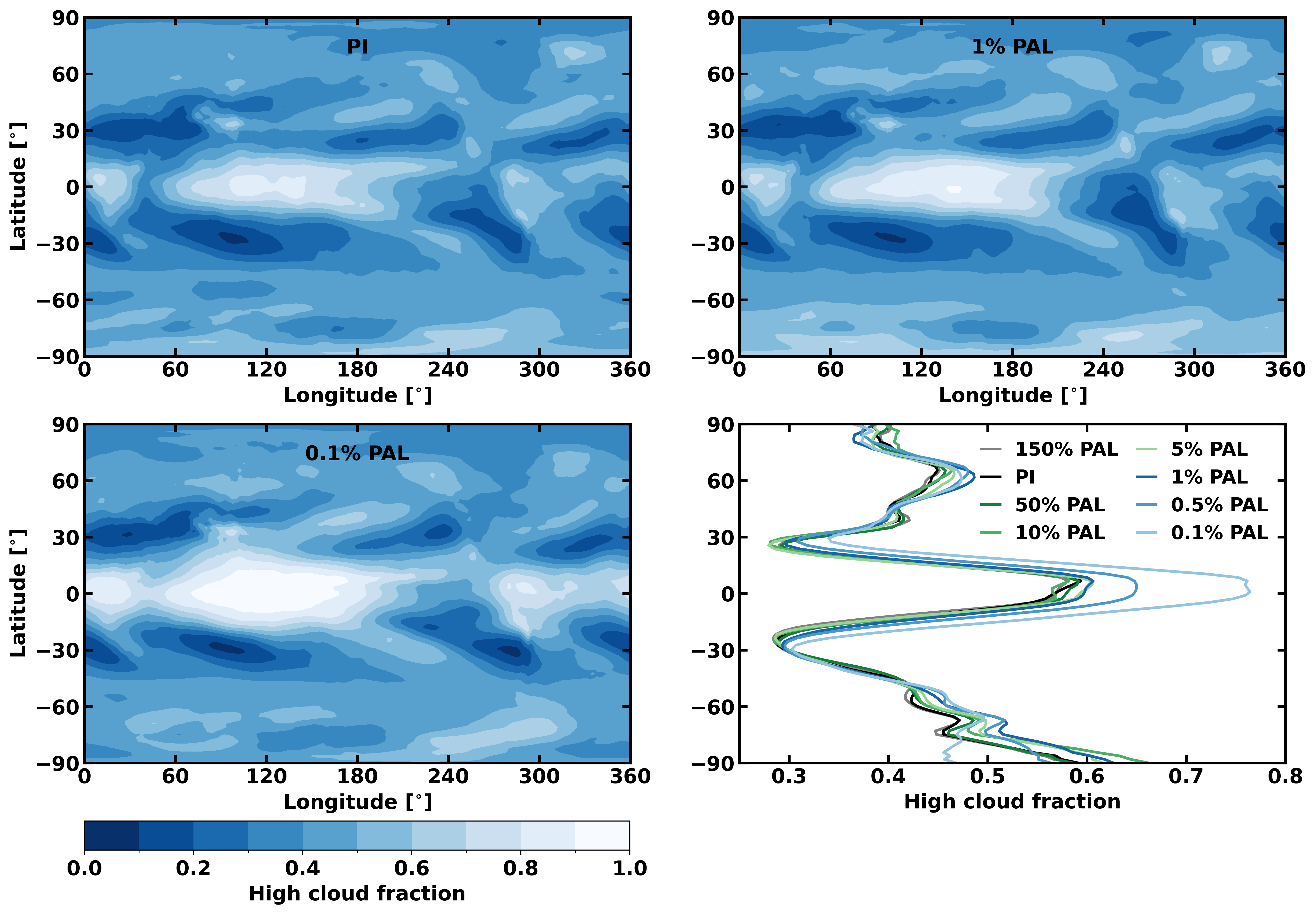}
    \caption{The \textbf{top left}, \textbf{top right}, and \textbf{bottom left} panels show the high cloud fraction as a function of longitude and latitude for the PI, 1\% PAL, and 0.1\% PAL scenarios, respectively. \textbf{Bottom right}: The high cloud fraction (high clouds are defined as clouds at pressures $<400$ hPa) is averaged over longitude and shown as a function of latitude for all the WACCM6 simulations. } 
    \label{High cloud fraction figure}
\end{figure}

The fractional coverage of high clouds (high clouds are defined as clouds at pressures $<400$ hPa and are therefore relevant to the TTL) versus latitude is shown in Fig.~\ref{High cloud fraction figure}. The coverage of high clouds for the PI, 1\% PAL, and 0.1\% PAL is also shown in the same figure. The O$_2$ mixing ratio affects high cloud coverage across the entire planet, but especially in the tropics. There is a $\approx30$\% increase in the peak high cloud fraction at the equator ($0^\circ$ latitude) from the 150\% PAL simulation to the 0.1\% PAL simulation, although the peak high cloud fraction for the simulations between the O$_2$ concentrations of 150\% PAL to 1\% PAL at the equator all fall within 6\% of each other. Low clouds (surface -- 700 hPa) and medium clouds (700 -- 400 hPa) exhibit very little deviation between the simulations.

\begin{figure*}[t!]
	\centering
	\includegraphics[width=1\textwidth]{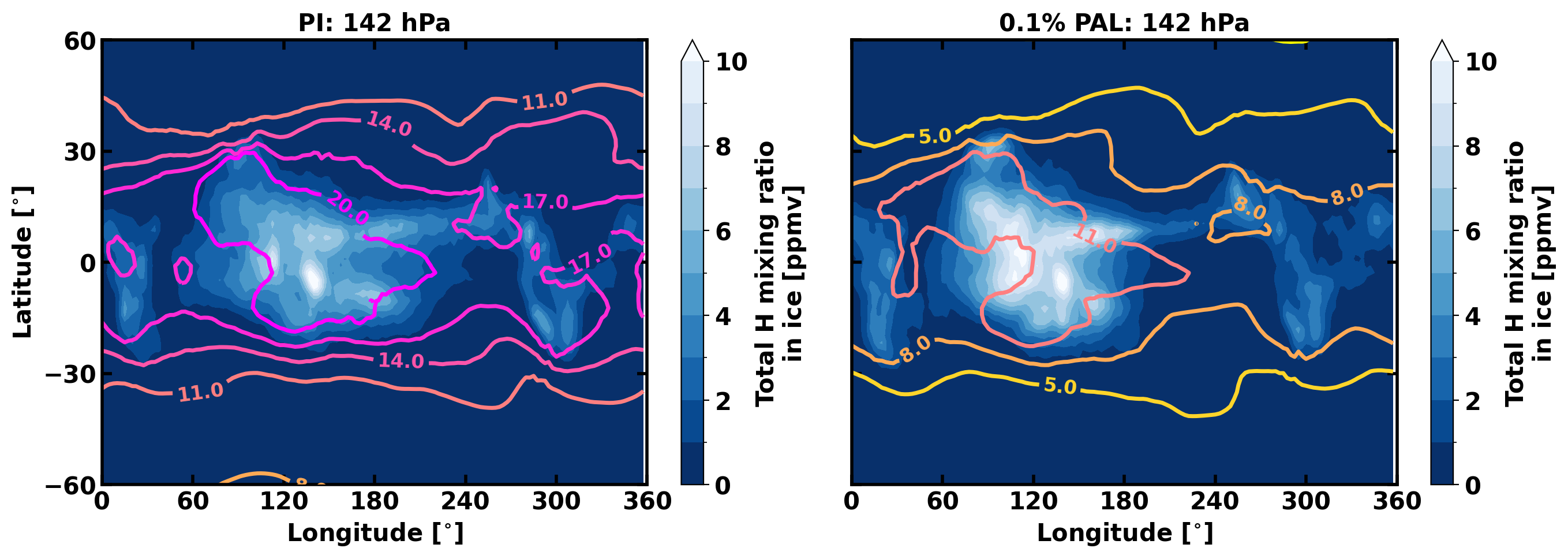}
    \caption{The atmospheric ice content (blue-white shading) and the amount of atmospheric H$_2$O (magenta-orange-yellow contours) is shown on the latitude-longitude grid of the Earth. Both are given in terms of ppmv. The PI simulation (\textbf{left}) and the 0.1\% PAL simulation are shown (\textbf{right}). Both scenarios are shown for a pressure level of 142 hPa, which corresponds to an altitude of $\approx 14$ km near the base of the TTL.} 
    \label{Water and Ice figure}
\end{figure*}

The total hydrogen mixing ratio bound in ice (ice water content and ice clouds) and H$_2$O for the PI and 0.1\% PAL cases is shown in Fig.~\ref{Water and Ice figure} between $-60^\circ$ to $+60^\circ$ latitude. At 142 hPa (roughly 14 km in altitude), the 0.1\% PAL scenario has a greater amount of ice content and a lower amount of H$_2$O. Thus, due to the freeze-drying of the atmosphere, fewer hydrogen atoms are able to contribute to the stratospheric total hydrogen mixing ratio, $f_\textrm{T}(\textrm{H})$, when compared to the PI case. Both simulations show a larger build-up of total ice water, ice clouds, and H$_2$O content over the western pacific (e.g., between $90$ -- $150^\circ$ East), as is found on modern Earth \citep{newell1981stratospheric, 1998GeoRL..25.4165D, 2020GeoRL..4786320D}.

% High cloud fraction figure

% Water and Ice figure

In the cases we have simulated, the interplay between UV radiation, O$_2$ mixing ratio, and 3D transport, is especially important for predicting the O$_3$ distribution. The amount of O$_3$ and the incoming UV affect temperatures around the tropopause, and ultimately how much H$_2$O can reach the stratosphere. Here we are exploring the impact of O$_2$ levels alone on the TTL temperatures and thus atmospheric escape. Other factors will have influenced TTL temperatures for the past 2.4 billion years, and these will be discussed in Section~\ref{Discussion}.

\subsection{Tropical tape recorder}

\begin{figure*}[t!]
	\centering	\includegraphics[width=1\textwidth]{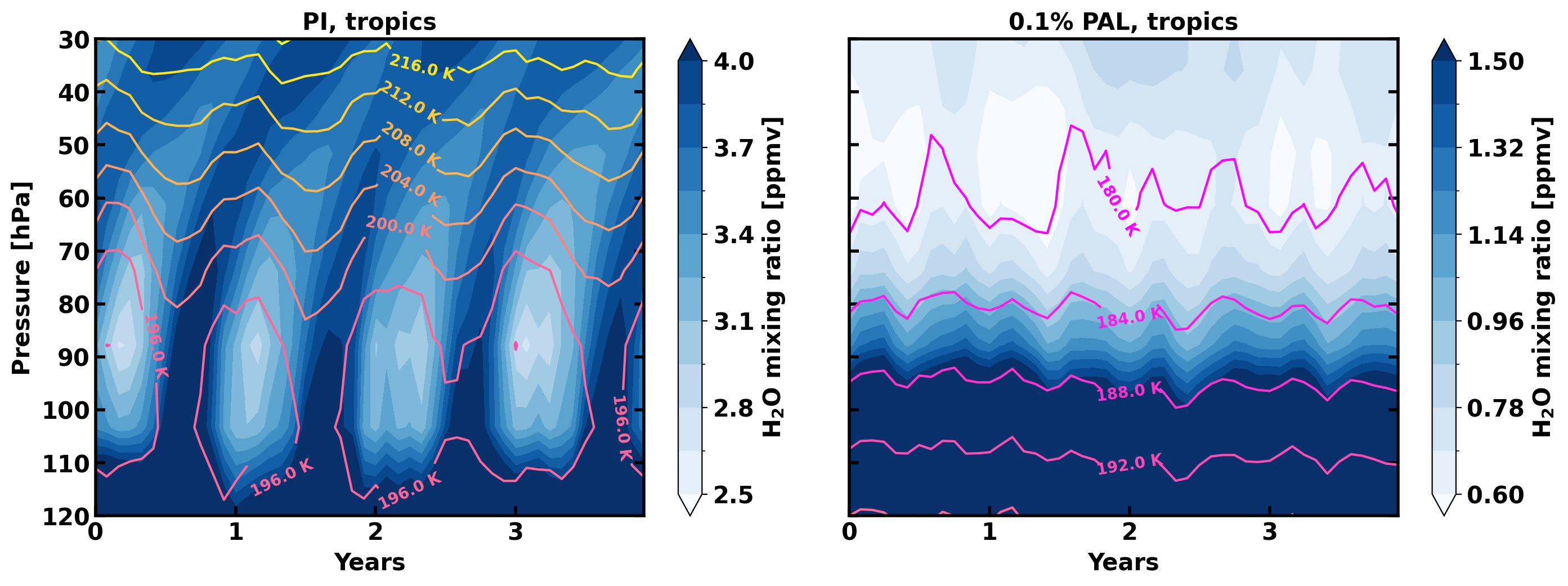}
    \caption{The water vapour mixing ratio in ppmv is shown for the PI scenario (\textbf{left}) and the 0.1\% PAL scenario (\textbf{right}) between pressures of 120 -- 30 hPa. Each panel shows four years in a row of each simulation, displaying the zonally averaged and latitudinally weighted H$_2$O mixing ratio in the tropics between $\pm 24^\circ$. Whites show the lowest mixing ratios and darker blues show progressively larger mixing ratios. Note the different scales on each colour bar. The contours show the atmospheric temperature in kelvin, with yellow indicating higher temperatures and magenta indicating lower temperatures.} 
    \label{Tape recorder changes figure}
\end{figure*}

The tropical ``tape recorder'' 
\citep{1995GeoRL..22.1093M, 2007GeoRL..34.9804L, 2017ACP....17.4337G} for H$_2$O is the phenomenon where seasonal temperature fluctuations in the TTL modulate the mixing ratio of H$_2$O that enters the stratosphere over an annual cycle. The temperature of the TTL itself is affected by several processes, including radiative heating and cooling, deep convection, sea surface temperatures, and the Quasi-Biennial Oscillation \citep{2009RvGeo..47.1004F, 2012ACP....1212183P, 2013JGRD..118.9658G, 2020GeoRL..4789533T}. Note that the tape recorder effect exists for some other molecules too, including CO \citep{2006GeoRL..3312811S, 2007GeoRL..34.9804L, 2008GeoRL..35.5801P}. The Brewer-Dobson circulation \citep{1926RSPSA.110..660D, 1949QJRMS..75..351B, 2014RvGeo..52..157B}, over a period of months, transports H$_2$O upward \citep{2017ACP....17.4337G}, causing the distinct tape recorder effect  \citep[WACCM6 is able to generate this effect - see the PI simulation in Fig.~\ref{Tape recorder changes figure} and also][]{2019JGRD..12412380G}. The seasonally-varying temperatures in the TTL imprint a record on the observed mixing ratio of H$_2$O which is altered in the middle atmosphere in the subsequent months.  

In the PI simulation, the tape recorder is easily visible over a 4-year time frame with a period of 1 year (see Fig.~\ref{Tape recorder changes figure}). For the PI case, the peak seasonal differences in the tape recorder effect are of order 1.5 ppmv at $\approx 90$ hPa. Note that this amplitude is much smaller than the tape recorder variation \cite{2023MNRAS.524.1491L} found for an Earth-like exoplanet with a higher eccentricity of 0.4.

At low oxygenation states ($< 1$\% PAL), seasonal differences in temperature and the Brewer-Dobson circulation are still present, but their magnitude is smaller so the tropical tape recorder effect is muted. Whilst the H$_2$O mixing ratios still vary, the tape recorder effect is no longer visible from the scale shown in Fig.~\ref{Tape recorder changes figure} at 0.1\% PAL because there is no clear periodic signal. Variance-preserving Fourier spectra show that the annual harmonic (1 cycle yr\textsuperscript{-1}) dominates the stratospheric H$_2$O variability in the PI simulation between 100 and 30 hPa. In contrast, the 0.1\% PAL simulation exhibits a much weaker annual contribution relative to total variance at these pressure levels, consistent with the absence of a tape-recorder signal. Therefore, the seasonal mechanism which transports more water vapour into the stratosphere during warmer periods in the TTL is no longer effective in the 0.1\% PAL case.

% Tape recorder changes figure

\subsection{Total hydrogen mixing ratio}

After H$_2$O passes through the cold trap, the first bottleneck for the total hydrogen mixing ratio of temperate Earth-like atmospheres has been passed. We now assess the quantitative effect of this bottleneck, and the relative contributions of each of the four major hydrogen bearing species in our simulations.

% Temp total H figure
\begin{figure*}[t!]
	\centering
	\includegraphics[width=1\textwidth]{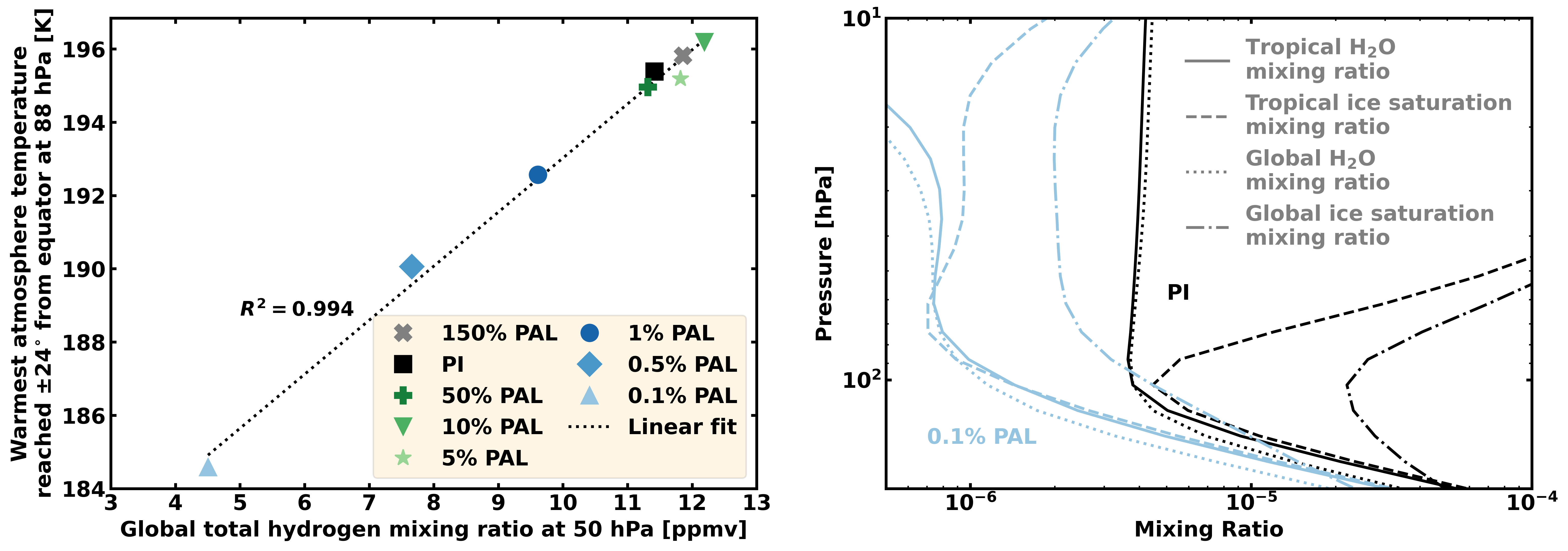}
    \caption{\textbf{Left}: The globally averaged total hydrogen mixing ratio ($f_\textrm{T}(\textrm{H})$) at 50 hPa is plotted against the warmest temperature reached within $\pm24^\circ$ (tropical latitudes) from the equator at 88 hPa for each simulation. Linear regression gives a coefficient of determination ($R^2$) of 0.994, which is showed by the dotted black line. The different O$_2$ mixing ratios are shown by various markers: 150\% PAL (grey cross), PI (black square), 50\% PAL (dark green plus), 10\% PAL (green downward triangle), 5\% PAL (light green star), 1\% PAL (dark blue circle), 0.5\% PAL (blue diamond), and 0.1\% PAL (light blue upward triangle). \textbf{Right}: The H$_2$O mixing ratio is plotted against pressure for the PI (black) and 0.1\% PAL (light blue) simulations, in terms of its tropical (unbroken line) and global (dotted line) average. Also shown is the ice saturation mixing ratio which depends on temperature and pressure and is shown for the tropical (dashed) and global (dash-dotted) average. The tropical average is for latitudes $\pm 24^\circ$ from the equator.}
    \label{Temp total H figure}
\end{figure*}

Fig.~\ref{Temp total H figure} displays the warmest atmospheric temperature reached in each simulation at a pressure of 88 hPa ($\approx 18$ km for Earth's atmosphere)\footnote{88 hPa is the pressure level in our simulations closest to the winter climatic tropopause of 86-88 hPa, although note that the summer climatic tropopause is closer to 100 hPa \citep{kim2012tropical, pilch2016tropical}} in the tropics (within $\pm24^\circ$ from the equator) against the globally averaged $f_\textrm{T}(\textrm{H})$ at the same pressure level. There is a strong positive correlation with a coefficient of determination of $R^2=0.994$ between these two variables. This freeze-drying relationship arises because the warmest temperatures in the tropical atmosphere effectively control the maximum amount of water vapor that can be transported upward through to the stratosphere. As a result, there is a weaker correlation present when comparing $f_\textrm{T}(\textrm{H})$ with the global mean temperature at the same pressure level, and worse correlation still when comparing to the warmest temperature found anywhere at 88 hPa. 

More importantly, Fig.~\ref{Temp total H figure} also demonstrates how it is the ice saturation vapor pressure \citep{murphy2005review} in the tropical atmosphere that acts to limit the transport of H$_2$O upwards. The tropical ice saturation mixing ratio curve is within a factor of 1.18 and 1.01 when compared to the globally averaged H$_2$O mixing ratio at 100 hPa ($\approx16$ km in altitude) for the PI and 0.1\% PAL atmospheres, respectively. When comparing the global ice saturation mixing ratio curve, the discrepancy is a factor of 5.8 and 3.9 times, respectively. This means that 1D models that are based on globally averaged temperatures may overestimate the stratospheric H$_2$O abundance for Earth-like oxygenated simulations by up to a factor of 6 (accounting for all simulations here, the discrepancy range is 3.8 -- 6.0). 

Fig.~\ref{Temp total H figure} suggests that the warmest TTL temperatures are the controlling factor in each atmospheric scenario, instead of the atmospheric composition. Yet because composition (i.e., the O$_2$ mixing ratio) is the variable that is altered in each scenario, the atmospheric composition is the controlling factor for the TTL temperatures. Hence, the oxygenation state of the atmosphere is indirectly controlling the upward flow of hydrogen atoms and affecting the diffusion-limited hydrogen escape rate, and this is what we refer to as the `oxygen valve'.

\begin{figure*}[t!]
	\centering
	\includegraphics[width=1\textwidth]{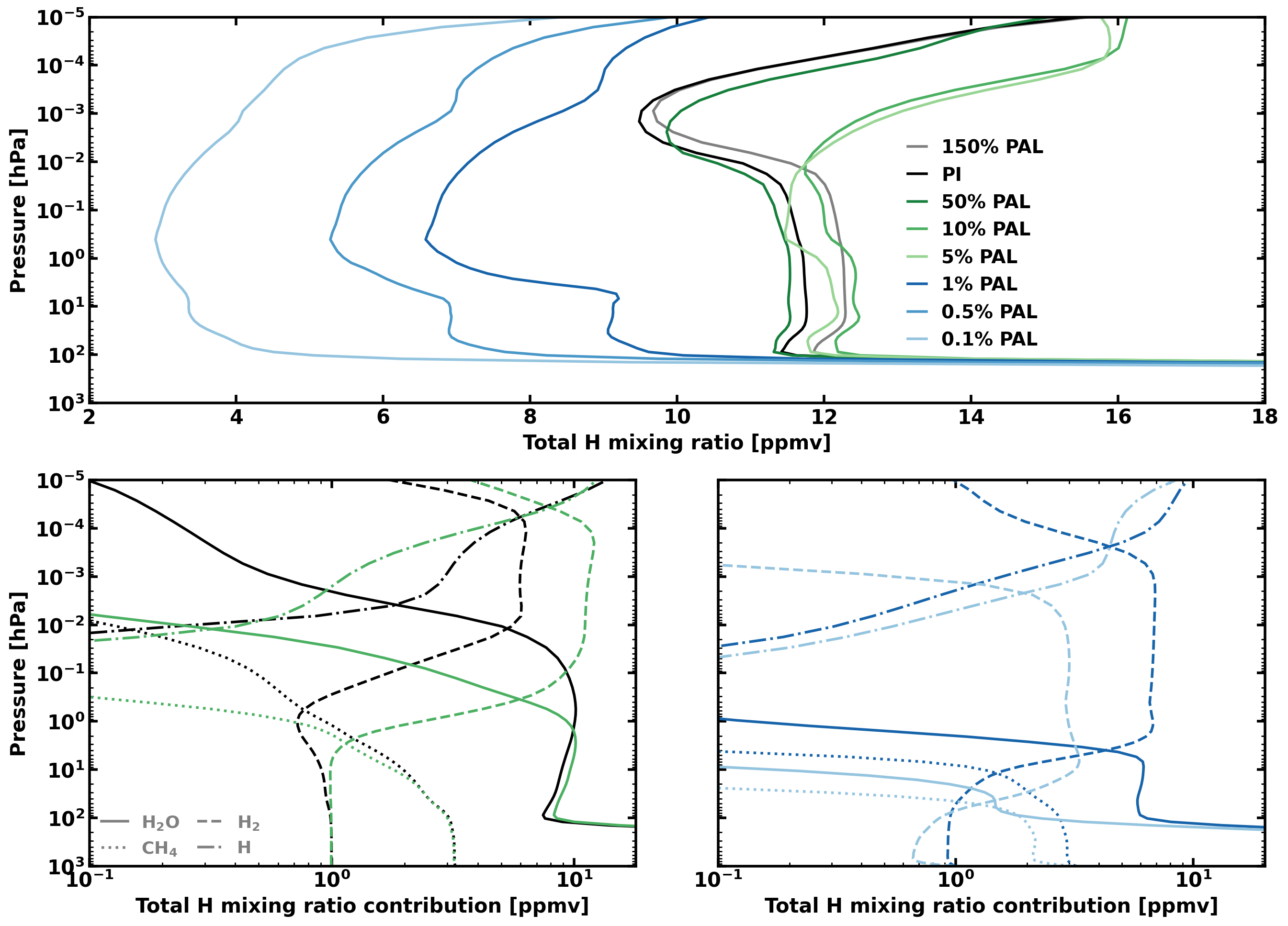}
    \caption{\textbf{Top}: $f_\textrm{T}(H)$, the total hydrogen mixing ratio is given in parts per million by volume (ppmv) and is shown against pressure for the 150\% PAL (grey), PI (black), 50\% PAL (dark green), 10\% PAL (green), 5\% PAL (light green), 1\% PAL (dark blue), 0.5\% PAL (blue), and 0.1\% PAL (light blue) simulations. $f_\textrm{T}(H)$ is calculated using Eq.~\eqref{Equation 2}. Hence, the mixing ratio contribution from H, H$_2$O, H$_2$, and CH$_4$ is multiplied by a factor of 1, 2, 2, and 4, respectively. \textbf{Bottom left}: The total hydrogen mixing ratio contribution from H (dash dotted), H$_2$ (dashed), H$_2$O (unbroken), and CH$_4$ (dotted) is shown for the PI and 10\% PAL simulations and \textbf{bottom right}: the 1\% PAL and 0.1\% PAL simulations.}
    \label{Total hydrogen figure}
\end{figure*}

Fig.~\ref{Total hydrogen figure} presents the globally averaged $f_\textrm{T}(\textrm{H})$ vertical profile throughout the atmosphere, which is calculated from Eq.~\ref{Equation 2} for all the simulations. The mixing ratio of total hydrogen in the stratosphere (at 1 hPa) is $f_\textrm{T}(\textrm{H}) = 11.7$ ppmv in the PI simulation. Turbulent mixing causes $f_\textrm{T}(\textrm{H})$ to remain roughly constant between the lower stratosphere and homopause. For the 10\% PAL, 1\% PAL, and 0.1\% PAL simulations, the stratospheric $f_\textrm{T}(\textrm{H})$ values at 1 hPa are approximately 12.4, 6.9, 3.0 ppmv, respectively.

Fig.~\ref{Total hydrogen figure} also shows the total hydrogen contribution from the four species that carry the majority of hydrogen atoms (H, H$_2$, H$_2$O, and CH$_4$). These panels show that for the PI simulation (black lines), H$_2$O is the primary carrier of hydrogen atoms in the lower atmosphere. Above the cold trap in the TTL, the H$_2$O mixing ratio increases until it reaches a maximum mesospheric value ($\approx 5$ ppmv) due to CH$_4$ reacting with OH. At a pressure of approximately 0.01 hPa, H$_2$ then becomes dominant. Photolysis of H$_2$O, CH$_4$, and H$_2$, as well as diffusive separation, cause atomic H to become the primary carrier at pressures less than $10^{-4}$ hPa. 

Decreasing levels of O$_2$ act to shift the pressures at which particular species are the principal hydrogen carrier. H$_2$ begins to increase in abundance at lower altitudes as the atmospheric O$_2$ concentration is decreased. For instance, with 1000 times less O$_2$ (0.1\% PAL case, light blue lines), H$_2$ is dominant at pressures less than 10 hPa, due to both H$_2$O condensation and photolysis at the higher pressure levels. Again, resulting from diffusive separation, H always dominates towards the top of the model in the lower thermosphere. CH$_4$ is increasingly lost in the troposphere due to enhanced H$_2$O photolysis and OH production, acting to reduce CH$_4$ lifetimes \citep[see the following papers for a more in depth discussion on CH$_4$ lifetimes:][]{2022RSOS....911165C, ji2023comparison, ji2024correlated}. This loss adds to tropospheric H$_2$O which is ultimately halted by the cold trap. If the CH$_4$ mixing ratio at the TTL is greater than half that of H$_2$O, then CH$_4$ is the primary contributor to hydrogen escape. However, CH$_4$ is never the main H carrier in the scenarios we present. As we will discuss later, it is unclear whether this would have been the case for some portions of the Proterozoic. 

% Total hydrogen figure all cases

\subsection{Diffusion-limited escape}

Now we have assessed $f_\textrm{T}(\textrm{H})$ throughout the atmosphere we can predict how the rate of diffusion-limited hydrogen escape varies between the different simulated atmospheres with WACCM6. We present the predicted escape rates for each simulation in Table~\ref{Escape rate table}.

\begin{table}
\caption{This table shows the eight simulations where the mixing ratio of O$_2$, given in terms of present atmospheric level (PAL), was varied. The total hydrogen mixing ratio, $f_\textrm{T}(\textrm{H})$, in ppmv at the homopause, which is taken to be at a pressure of $10^{-3}$ hPa. The diffusion-limited hydrogen escape rate is given by $\Phi_\textrm{esc}$ in mol yr\textsuperscript{-1} and compared to the predicted escape rate in the PI simulation ($\Phi_\textrm{esc,PI}$).}
\centering
\begin{tabular}{ccccc}
%\toprule
Simulation &  O$_2$ [PAL] & Homopause $f_\textrm{T}(\textrm{H})$ [ppmv] & $\Phi_\textrm{esc}$ [mol yr\textsuperscript{-1}] & $\Phi_\textrm{esc}$/$\Phi_\textrm{esc,PI}$\\ \hline 
150\% PAL & 1.500 & 9.68 & $7.71\times10^{10}$ & 1.02 \\
PI  & 1.000 & 9.52 & $7.54\times10^{10}$ & 1.00 \\
50\% PAL & 0.500 & 10.05 & $7.79\times10^{10}$ & 1.03 \\
10\% PAL & 0.100 & 12.74 & $9.04\times10^{10}$ & 1.20 \\
5\% PAL &  0.050 & 13.07 & $8.95\times10^{10}$ & 1.19 \\
1\% PAL &  0.010 & 8.45 & $5.11\times10^{10}$ & 0.68 \\
0.5\% PAL & 0.005 & 6.92 & $3.89\times10^{10}$ & 0.57 \\
0.1\% PAL & 0.001 & 4.09 & $1.94\times10^{10}$ & 0.26 \\
\end{tabular}
\label{Escape rate table}
\end{table}

The $f_\textrm{T}(\textrm{H})$ profile shows that for the cases with $\leq$ 1\% PAL of O$_2$, the rate of hydrogen escape is lower than that found in the PI case ($\Phi_\textrm{esc,PI}$). The rates for the 0.1\% PAL, 0.5\% PAL, and 1\% PAL simulations are: $0.26\ \Phi_\textrm{esc,PI}$, $0.57\ \Phi_\textrm{esc,PI}$, and $0.68\ \Phi_\textrm{esc,PI}$, respectively. We also predict that the escape rate could have been slightly higher when O$_2$ levels were both greater (e.g. 150\% PAL) and lesser (e.g. 5 -- 10\% PAL) than the present day value, with the maximum predicted hydrogen escape rate in the 10\% PAL simulation at $1.20\ \Phi_\textrm{esc,PI}$. This shows that the TTL heating and predicted escape rate has a non-linear behavior with O$_2$ concentration. Following the Great Oxidation Event, Earth's O$_2$ levels are thought to have fluctuated between 0.1\% PAL and 150\% PAL, although the precise timings and magnitude of these fluctuations are debated \citep{2020SciA....6.1420C, 2020PreR..34305722S, 2020AsBio..20..628P, yierpan2020recycled, krause2022extreme, 2023AREPS..51..253M, 2023GSAB..135..753X, 2024NatCo..15.6794F, 2024NatGe..17..667S}, with these fluctuations being due to various physical factors, such as carbonate amassing in the crust \citep{2024NatGe..17..458A}, and large igneous province volcanism combined with weathering \citep{2024ComEE...5..609L}.

\section{Discussion}
\label{Discussion}

When the Great Oxidation Event occurred approximately 2.4 billion years ago, O$_2$ increased by several orders of magnitude in the atmosphere in a time frame of approximately 200 Myr \citep{2017PNAS..114.1811G}. For the first time, these higher levels of O$_2$ resulted in the generation of a significant O$_3$ layer which may have created a stratospheric thermal inversion \citep{kasting1985oxidant, 2003AsBio...3..689S, 2022RSOS....911165C}. The O$_3$ layer provided an additional shield for Earth's surface from ultraviolet radiation. However, the exact thickness of the O$_3$ layer over geological time is difficult to determine. Many uncertainties exist upon the atmospheric pressure and composition throughout the Proterozoic \citep{2014Natur.506..307L, 2020PreR..34305722S, 2020SciA....6.1420C}, as well as there being differences in model predictions under similar assumptions \citep{2017ApJS..231...12W, 2022RSOS....911165C, 2022CliPa..18.2421Y, 2023RSOS...1030056J, ji2024correlated}. With an oxygenated biosphere, the conditions were set for the evolution of oxygen-dependent animals \citep{2020Gbio...18..260C} and the eventual colonization of land \citep{dunn2013evolution}. Additionally, the GOE may have coincided with the end of relatively high levels of hydrogen escape \citep{2013ChGeo.362...26Z, 2019GeCoA.244...56Z}. Whether other terrestrial worlds go through a similar path of geological and atmospheric evolution is unknown, although it has been modelled under different conditions and assumptions \citep{2018AsBio..18..856G, 2018haex.bookE.189O, 2021NatGe..14..138O, 2021AGUA....200294K, 2022ApJ...933..115K}.

Our work represents an initial step towards a quantitative estimate for hydrogen escape since the Proterozoic began 2.4 billion years ago. Overall, our results suggest that hydrogen escape during the Proterozoic could have been between 3.8 times lower and up to 1.2 times greater than the calculated pre-industrial escape rate when accounting for O$_2$ changes only. The present day loss of hydrogen is insignificant on geological timescales, such that the fluctuations in O$_2$ from the start of the Proterozoic to the present day likely did not in itself cause substantial levels of hydrogen escape. In what follows, we discuss how our simulations can be used as a platform for future work towards the goal of reconstructing Earth's past atmospheric hydrogen escape.

\subsection{The oxygen-ozone valve}
\label{The oxygen-ozone valve section}

When only changing the atmospheric O$_2$ (and N$_2$) mixing ratio, as O$_2$ increases from 0.1\% PAL to 150\% PAL, the total atmospheric O$_3$ column increases \citep{2022RSOS....911165C}. Note that increasing quantities of total O$_3$ does not always lead to an increase in temperatures in the TTL, because the location of the O$_3$ abundance peak in the atmosphere, and the magnitude of the peak, can move with O$_2$ mixing ratio. This effect is illustrated by the greater total hydrogen mixing ratio at the homopause in the 10\% and 5\% PAL cases when compared to the PI case (e.g., see Fig.~\ref{Zonal mean temperature figure}, Fig.~\ref{Temp total H figure}, and Fig.~\ref{Total hydrogen figure}).

Our work shows that O$_2$ levels during the Proterozoic may have partially controlled the diffusion-limited flux for hydrogen escape. However, this will depend on several other factors, such as the continental distribution and the Earth's obliquity which can affect atmospheric dynamics and seasonal cycles, respectively. The fainter Sun, the albedo of the Earth through time, and various greenhouse gases (including CH$_4$) which can act to cool the stratosphere but warm the surface \citep{2024PNAS..12119228L}, will have also impacted TTL temperatures and hydrogen escape rates. For instance, a cooling of the stratosphere can increase the amount of O$_3$, just like the current increasing atmospheric content of CO$_2$ is aiding a possible, albeit complex, recovery of the O$_3$ layer \citep{2002JGRD..107.4049R, fahey2018scientific, 2022WCD.....3..139I, 2021ACP....2111041M}.

There exist estimates of many of these aforementioned variables through Earth's history, but each of them come with uncertainties and vary throughout the Proterozoic. \cite{liu2025evolution} suggest that the O$_3$ layer could have been kept at low levels due to protracted atmospheric iodine concentrations throughout the Proterozoic. Atmospheric iodine would destroy O$_3$ through catalytic cycles (WACCM6 does not currently include iodine in its chemical mechanism). The computed WACCM6 O$_3$ columns are lower than the baseline estimates from \cite{liu2025evolution}, but when \cite{liu2025evolution} included substantial atmospheric iodine (e.g. $20\times$ modern marine iodine concentrations that was emitted to the atmosphere), then their O$_3$ columns are lower than the WACCM6 results. If we incorporated iodine, then this could lead to a reduced rate of predicted hydrogen escape at a given O$_2$ concentration due to diminished O$_3$ heating. 

\subsection{Model limitations} 
\label{Model limitations section}

Our atmospheric predictions are relevant for both Earth's past and future climate states. However, this statement comes with several caveats. The moist-adiabatic lapse rate (the rate at which the temperature of a parcel of saturated air decreases with altitude as it rises in the atmosphere under adiabatic conditions), strongly depends on temperature. Thus, the total hydrogen mixing ratio in the troposphere of the Earth is determined by the incoming radiation, albedo, and greenhouse effect. For modern Earth, these parameters are relatively well known. We did not attempt to fully account for all climatic changes since the Proterozoic began; instead, as a first step, we focused on the effect on the escape rate by changing O$_2$ only. The range of O$_2$ concentrations we accounted for are relevant to the past 2.4 billion years and are not relevant to the Archean eon (4 -- 2.4 Gyr ago) when molecular oxygen was much lower than present day and Proterozoic (2.4 -- 0.541 Gyr ago) concentrations \citep{2020SciA....6.1420C}. We remind the reader that WACCM6 is designed for the modern Earth and the current set of simulations do not account for a faster rotation rate in the past \citep{1987PreR...37...95Z, 2016GeoRL..43.5716B, laskar2023did} or different continental coverage \citep{2021NRvEE...2..358M}. Additionally, we note that changes to the orbital parameters, rotation rate, or obliquity, may affect the tape recorder signal.

Since \cite{2022RSOS....911165C} was published, \cite{ji2023comparison} showed that absorption of light by H$_2$O and CO$_2$ in the S-R bands, as well as scattering in the S-R bands, are increasingly important at lower atmospheric O$_2$ mixing ratios. The simulations presented in this work do not include either of these treatments. Some WACCM6 simulations (at 10\% PAL, 1\% PAL, and 0.1\% PAL) with H$_2$O and CO$_2$ absorption in the S-R bands were included in \cite{ji2023comparison}, and we also ran some simulations with fixed surface fluxes (e.g. for CH$_4$). Whilst the absolute mixing ratios do deviate when this absorption is included, the effect of changing O$_2$ on temperature is stronger. In terms of scattering, \cite{ji2023comparison} showed that it is most important at $\leq1$\% PAL of O$_2$. Inclusion of scattering will decrease the total O$_3$ column and likely result in a cooler TTL region. Again, the conclusion that hydrogen escape can be be modified by O$_2$ concentrations remains unaffected, especially when the prescribed CH$_4$ mixing ratio sets a minimum on the hydrogen escape rate because it is not cold-trapped. Nevertheless, future simulations where WACCM6 includes both absorption and scattering in the S-R bands will provide more comprehensive atmospheric calculations.

\subsection{Future work for Earth} 

The different climate states that might have existed throughout the Proterozoic and Phanerozoic\footnote{By climate states, here we refer to the variety of factors that would have affected weather, atmospheric circulation systems, and surface temperature. These include glaciations and hotter periods, as well as the various continental land masses, fainter Sun, and greenhouse gas concentrations.} (0.541 Gyr ago -- present day) should be simulated in future work. A hotter troposphere may have existed due to increased amounts of CH$_4$ or CO$_2$, or both. In this case, the warmer TTL would have increased hydrogen escape rates by elevating the amount of H$_2$O reaching the stratosphere. On the other hand, colder temperatures during Earth's glacial periods would cool the tropopause \citep{2019JGRD..12411819G} and reduce the stratospheric H$_2$O concentration.

Even with a much colder TTL, CH$_4$ bypases the control of the oxygen valve or a colder troposphere because CH$_4$ does not condense in Earth's atmosphere like H$_2$O. Thus, the Proterozoic and Phanerozoic hydrogen escape rate could have instead been controlled by the CH$_4$ abundance rather than the O$_2$ abundance, assuming that its mixing ratio reaching the TTL was greater than half the mixing ratio of H$_2$O in the TTL - see Eq.~\ref{Equation 2}. The current problem with evaluating this effect is that estimates for Proterozoic CH$_4$ concentrations vary by several orders of magnitude \citep{2003Geo....31...87P, 2016E&PSL.434...42D, 2018Geo....46..139Z, 2018haex.bookE.189O, 2019E&PSL.522...48L, 2020SciA....6.1420C, cadeau2020carbon} and may be greater or lower than the present day mixing ratio. The thinner simulated O$_3$ columns found from 3D modeling \citep{2022RSOS....911165C, 2022CliPa..18.2421Y} and updated 1D calculations \citep{ji2024correlated}, as well as the inclusion of iodine chemistry \citep{liu2025evolution}, suggest that more UV radiation would have penetrated deeper into the atmosphere for much of the Proterozoic and potentially resulted in lower CH$_4$ lifetimes at O$_2$ concentrations of $\leq 1$\% PAL. In this situation, greater surface-to-atmosphere fluxes of CH$_4$ would have been required to sustain current $\sim1$ ppmv mixing ratios \citep[such fluxes may not be plausible;][]{daines2016effect, Olson2016LimitedGreenhouse, 2019E&PSL.522...48L}, or the higher estimates that some studies suggest in the region of $\sim10$ -- $100$ ppmv \citep{2003Geo....31...87P, 2017Geo....45..231F, zhao2018terrestrial, fakhraee2019proterozoic}. Closer in time to the present day, CH$_4$ mixing ratios in the Phanerozoic (0.51 Gyr ago - present day) have been simulated to be between 0.1 -- 12 ppmv \citep{beerling2009methane}.

Clearly, there is much to still learn about why the different models predict specific atmospheric abundances for molecules such as O$_3$, which has significant horizontal and vertical variation \citep{2022RSOS....911165C, 2022CliPa..18.2421Y}, and CH$_4$ \citep{kasting2025evolution}. The entire conceivable parameter space has certainly not been fully explored; such endeavors utilising 3D models are welcomed but will likely prove computationally expensive. Attempts to define the CH$_4$ concentrations in the various geological eons are an important part of the history of hydrogen escape, but for now, the mixing ratio of CH$_4$ through time remains poorly constrained. Ascertaining its production rate in the Proterozoic will help to establish its concentrations in that eon \citep{kasting2025atmospheric}.

Eventually, as the Sun's luminosity increases, the Earth will heat up, CO$_2$ will be sequestered out of the atmosphere through carbonate-silicate weathering, and it is predicted that the Earth will lose all of its water through the moist greenhouse effect approximately 2 Gyr in the future \citep{1984Icar...57..335K, 1988Icar...74..472K, 2014GeoRL..41..167W, 2021NatGe..14..138O}, although the estimate of the time varies. But before that event occurs, previous studies have predicted that the lack of CO$_2$ will have a detrimental effect on the biosphere because plants are unable to survive at very low CO$_2$ concentrations \citep{1982Natur.296..561L, 1992Natur.360..721C}. 

Several calculations of the lifespan of the Earth's biosphere have been attempted \citep{1982Natur.296..561L, 1992Natur.360..721C, 2000TellB..52...94F, von2003biogenic, 2018AsBio..18..469R, 2021NatGe..14..138O, 2023IJAsB..22..272M, Graham_2024_Extension}. For instance, \cite{2021NatGe..14..138O} calculated that in $1.08 \pm 0.14$ billion years ($1\ \sigma$), Earth's atmospheric O$_2$ will drop to 1\% PAL. More recently, \cite{Graham_2024_Extension} suggested that Earth's biosphere may last for up to 1.6 -- 1.86 Gyr. Most of these estimates for biospheric termination are before the Earth is predicted to lose its water and become, by the most basic definition, uninhabitable. As O$_2$ levels drop following the extinction of photosynthesizing life \citep{2021NatGe..14..138O}, O$_3$ will decrease, and the upper troposphere and TTL may be colder than under otherwise equivalent conditions, diminishing the amount of hydrogen available in the upper atmosphere which can escape to space and cause irreversible water loss (the future rates of hydrogen escape would likely have a negligible effect on oxidizing the Earth to reverse the loss of O$_2$ from ceasing photosynthesis). Hence, our results suggest that the ocean loss timescale could be extended. Thus, investigations aiming to assess the lifetime of Earth's future habitability should incorporate a 3D chemistry-climate model that includes comprehensive atmospheric oxygen chemistry and that can adequately represent the 3D movement of air parcels through the TTL. Additionally, it is possible that our WACCM6 simulations could be used as input to a model which has a higher top (lower minimum pressure), such as the WACCM-X model \citep{2018JAMES..10..381L}, or even used as an input into a computational model that explicitly calculates atmospheric escape \citep[e.g.,][]{2018PNAS..115..260D, 2023MNRAS.523..286S}.

\subsection{Future work for exoplanets}

Other planets outside of our solar system (exoplanets) may be habitable and evolve life. Whether they can keep liquid water on their surface is crucial for the persistence of habitability as it is currently defined.

The H$_2$O tropical tape recorder is a stratospheric signal of seasonality, impressing the varying TTL temperatures on the stratospheric H$_2$O mixing ratio. Prior work showed how a 15\% reduction in global O$_3$ \citep{Xie_2008_O3_tape} could cool the TTL and raise it in height, ultimately weakening the tape recorder signal. Here we find this effect exacerbated further as O$_2$ and O$_3$ reduce by orders of magnitude. Transmission spectra of exoplanet atmospheres probe the stratosphere, and whilst beyond the detection limit of today's telescopes, such chemical variability may be detectable with future technology or where exoplanets exhibit higher levels of variation that Earth \citep[e.g., for exoplanets with eccentricity $\geq0.4$][]{2023MNRAS.524.1491L} or different types of stratospheric variability \citep{Cohen_2022_LASO}. Seasonality of biosignatures and other gases \citep{Olson_2018_season}, combined with the H$_2$O tape recorder signal, may aid in determining whether a planet is inhabited. Should there be no tape recorder signal due to a tropopause that does not significantly vary with seasons, then this variability will be more difficult to observe, and H$_2$O will be harder to constrain due to a dry middle-upper atmosphere.

Simulations with other 3D models such as the LMD-g \citep{2022CliPa..18.2421Y} or the Unified Model \citep{2022MNRAS.517.2383B}, or those at higher resolutions \citep[e.g., LFRic-Atmosphere;][]{2023GMD....16.5601S}, may yield different quantitative results due to varying treatments of convection and water vapour microphysics, as well as alternative radiative transfer schemes. The stratospheric thermal inversion in Earth's atmosphere is due to O$_3$ which provides sufficient atmospheric heating when O$_2$ is present in quantities $\gtrsim 0.1$\% PAL \citep{2017ApJS..231...12W, 2022RSOS....911165C}. But a terrestrial exoplanet may not require O$_2$ to modulate the thermal structure: other atmospheric compositions can result in thermal inversions and may similarly act to enhance or reduce water loss. For example, Titan has a thermal inversion due to the prescence of hazes which absorb shortwave radiation \citep{1973Icar...20..437D, 2017aeil.book.....C}, and CN on hot super-Earths can also cause a thermal inversion \citep{2021MNRAS.500.2197Z}. Additionally, the heating rates will change depending on the distribution of stellar radiation over wavelengths: O$_3$ radiative heating significantly affects planets around F and G stars but less so around M and K stars \citep{2015P&SS..111...62G, 2022A&A...665A.156K, 2024MNRAS.531.1471D}. O$_2$ can also build up on exoplanets without the presence of life under specific conditions due to large amounts of CO$_2$  \citep{2007A&A...472..665S} or H$_2$O photolysis  \citep{2014ApJ...785L..20W}.

There are various planetary system parameters which affect the climate state of exoplanets. Crucial to the atmospheric dynamics is the planetary rotation rate and whether it is tidally locked \citep{2014RSPTA.37230084F, 2016A&A...596A.112T, 2017A&A...601A.120B, 2019AsBio..19...99D, 2025PSJ.....6....5B}. The circulation affects the atmospheric temperature structure and how tracers such as H$_2$O are advected. Other salient factors include the strength and spectral shape of the incoming radiation \citep{2020A&A...639A..99E}, the continental distribution or lack thereof \citep{2018ApJ...854..171L, 2021ApJ...910L...8Z, 2022MNRAS.513.2761M}, obliquity \citep{2019ApJ...877L...6K}, eccentricity \citep{2023MNRAS.524.1491L}, flares \citep{2021NatAs...5..298C,2023MNRAS.518.2472R}, simulation resolution \citep{2020ApJ...894...84S, 2021ApJ...913..101L, 2022ApJ...934..149S, 2022ApJ...940...87K, 2023NatAs...7.1070Y, Sergeev_2024_Impact, 2024ApJ...965....5G}, as well as the planetary Bond albedo and surface composition \citep{2011ApJ...729...54C, 2019ApJ...884...75D, 2020MNRAS.495....1M}. Ocean salinity can also be important in modulating the climate of rocky planets \citep{2022GeoRL..4995748O}.

By connecting exoplanet simulations to a model that includes the ionosphere (e.g., WACCM-X), one could attempt to answer interesting questions regarding the upper atmospheres of terrestrial exoplanets. For instance, how do the ionospheres of various exoplanets compare to Earth? Just like flares disturb the atmospheres of terrestrial planets \citep{2010AsBio..10..751S, 2019AsBio..19...64T, 2021NatAs...5..298C, 2023MNRAS.518.2472R}, do flares (especially for active M dwarf stars) impact their ionospheres differently from those that strike the Earth \citep{2009Ge&Ae..49..983L, 2018SpWea..16.1363S,2021SoPh..296..157H}? And could those differences manifest in a way which is detectable with future instruments \citep{2018NatAs...2..287M, 2019A&G....60a1.25M}?

The maximum difference we found in the diffusion-limited hydrogen escape between alternative oxygenation states when all other initial conditions and boundary conditions are kept constant is a factor of 4.7. Whilst this change is not significant for Earth under the assumptions made in this work, it potentially could be significant when the ocean loss timescale for a habitable terrestrial exoplanet is similar to a time that a planet could theoretically spend in the habitable zone (i.e. a planet with a 10 Gyr ocean loss timescale with a different oxygenation state could instead have an ocean loss timescale of $\approx 2$ Gyr). 

\section{Conclusions}
Since the dawn of the Proterozoic (2.4 Gyr ago), several properties of the Earth have changed. The continental configuration has shifted several times, the Sun's luminosity has increased, and the concentration of atmospheric gases such as O$_2$, CO$_2$, and CH$_4$ have fluctuated. 

This work used WACCM6 simulations of the Earth to explore the effect that the changing O$_2$ mixing ratio had on the total hydrogen mixing ratio at the homopause in order to predict the possible diffusion-limited hydrogen escape rate through geological time. A 3D model with chemistry and a global representation of water vapour microphysics is critical to simulating the transport of H$_2$O into the stratosphere and thus estimating diffusion-limited hydrogen escape when other atmospheric constituents (e.g., CH$_4$ and H$_2$, which we held constant at the surface) are not the dominant hydrogen bearing species. In our simulated scenarios we showed that 1D models may predict a stratospheric water vapour discrepancy of between a factor of 3.8 -- 6.0 depending on the O$_2$ mixing ratio when compared to 3D calculations.

We found that the atmospheric O$_2$ mixing ratio acts as a nonlinear valve on the total hydrogen mixing ratio in the stratosphere. The O$_3$ mixing ratio (which is affected by the abundance of O$_2$) around the tropical tropopause layer (TTL) determines the amount of UV heating that warms the TTL and controls the H$_2$O mixing ratio entering the stratosphere. The obliquity of Earth induces seasonal cycles, which affect phenomena such as the Brewer-Dobson circulation and the tropical tape recorder. We showed that the tropical tape recorder disappeared at 0.1\% PAL, which suggests that the present-day seasonal transport of H$_2$O molecules to the stratosphere may not have taken place at low O$_2$ mixing ratios.

Between 0.1\% PAL and 150\% PAL of O$_2$, the upward diffusion of H$_2$O is modulated in a non-linear manner, with 5\% and 10\% PAL having the maximum predicted diffusion-limited escape rate of hydrogen due to O$_3$ concentrations causing localized heating. The predicted escape rate changes by up to a factor of 4.7 between all of the simulations, meaning that whilst the oxygen valve affects diffusion-limited hydrogen escape, in the absence of other factors, it would not have resulted in relatively high levels of hydrogen escape and thus water loss since the GOE. 

In order to dissect Earth's atmospheric history, future work should investigate multiple other physical, chemical, and biological factors that may affect the hydrogen escape rate whilst also accounting for the oxygen valve we presented.

\codeavailability{WACCM6 is a publicly available code. The specific release used in this paper was CESM2.1.3, which can be downloaded from the following \href{https://escomp.github.io/CESM/versions/cesm2.1/html/downloading_cesm}{CESM downloads} page. Our code modifications are detailed here in the \href{https://github.com/exo-cesm/CESM2.1.3/tree/main/O2_Earth_analogues}{ExoCESM GitHub}.}

\dataavailability{The time-averaged data is currently available in the \href{https://doi.org/10.5061/dryad.ncjsxksvn}{Dryad data repository}.} %% use this section when having only data sets available

%\codedataavailability{} %% use this section when having data sets and software code available

%\sampleavailability{TEXT} %% use this section when having geoscientific samples available

%\videosupplement{TEXT} %% use this section when having video supplements available

%\appendix
%\section{}    %% Appendix A

%\subsection{}     %% Appendix A1, A2, etc.

%\noappendix       %% use this to mark the end of the appendix section. Otherwise the figures might be numbered incorrectly (e.g. 10 instead of 1).

%% Regarding figures and tables in appendices, the following two options are possible depending on your general handling of figures and tables in the manuscript environment:

%% Option 1: If you sorted all figures and tables into the sections of the text, please also sort the appendix figures and appendix tables into the respective appendix sections.
%% They will be correctly named automatically.

%% Option 2: If you put all figures after the reference list, please insert appendix tables and figures after the normal tables and figures.
%% To rename them correctly to A1, A2, etc., please add the following commands in front of them:

%\appendixfigures  %% needs to be added in front of appendix figures

%\appendixtables   %% needs to be added in front of appendix tables

%% Please add \clearpage between each table and/or figure. Further guidelines on figures and tables can be found below.

\authorcontribution{GJC set up and ran the simulations, performed the analysis, wrote the analysis and plotting code, and wrote the manuscript. DRM and CW supervised the project and provided comments on the manuscript. FSM and MB provided comments on the manuscript.} %% this section is mandatory

\competinginterests{The authors declare no competing interests} %% this section is mandatory even if you declare that no competing interests are present

%\disclaimer{TEXT} %% optional section

\begin{acknowledgements}
We would like to thank the reviewers for their comments which helped to improve the manuscript. G.J.C. thanks the Science and Technology Facilities Council for financial support during the PhD when the simulations were conducted (grant number ST/T506230/1). G.J.C. thanks James Rogers and Oli Shorttle for helpful discussions regarding Earth's history and hydrogen escape. F. Sainsbury-Martinez and C. Walsh would like to thank UK Research and Innovation for support under grant number MR/Z00019X/1 Additionally, C. Walsh would like to thank the University of Leeds and the Science and Technology Facilities Council for financial support (ST/X001016/1). M.B. appreciates support from a CSH Postdoctoral Fellowship. This work was undertaken on ARC4, part of the High Performance Computing facilities at the University of Leeds, UK.  The CESM project is supported primarily by the U.S. National Science Foundation.

\end{acknowledgements}

%% REFERENCES

%% The reference list is compiled as follows:

%% Since the Copernicus LaTeX package includes the BibTeX style file copernicus.bst,
%% authors experienced with BibTeX only have to include the following two lines:
%%
\bibliographystyle{copernicus}
\bibliography{example.bib}

@ARTICLE{Cohen_2022_LASO,
       author = {{Cohen}, Maureen and {Bollasina}, Massimo A. and {Palmer}, Paul I. and {Sergeev}, Denis E. and {Boutle}, Ian A. and {Mayne}, Nathan J. and {Manners}, James},
        title = "{Longitudinally Asymmetric Stratospheric Oscillation on a Tidally Locked Exoplanet}",
      journal = {\apj},
         year = 2022,
        month = may,
       volume = {930},
       number = {2},
          eid = {152},
        pages = {152},
          doi = {10.3847/1538-4357/ac625d},
archivePrefix = {arXiv},
       eprint = {2111.11281},
 primaryClass = {astro-ph.EP},
       adsurl = {https://ui.adsabs.harvard.edu/abs/2022ApJ...930..152C}
}

@ARTICLE{Olson_2018_season,
       author = {{Olson}, Stephanie L. and {Schwieterman}, Edward W. and {Reinhard}, Christopher T. and {Ridgwell}, Andy and {Kane}, Stephen R. and {Meadows}, Victoria S. and {Lyons}, Timothy W.},
        title = "{Atmospheric Seasonality as an Exoplanet Biosignature}",
      journal = {\apjl},
         year = 2018,
        month = may,
       volume = {858},
       number = {2},
          eid = {L14},
        pages = {L14},
          doi = {10.3847/2041-8213/aac171},
archivePrefix = {arXiv},
       eprint = {1806.04592},
 primaryClass = {astro-ph.EP},
       adsurl = {https://ui.adsabs.harvard.edu/abs/2018ApJ...858L..14O}
}

@ARTICLE{Zhou_2022_Ozone,
       author = {{Zhou}, Lingyu and {Xia}, Yan and {Zhao}, Chuanfeng},
        title = "{Influence of Stratospheric Ozone Changes on Stratospheric Temperature Trends in Recent Decades}",
      journal = {Remote Sensing},
         year = 2022,
        month = oct,
       volume = {14},
       number = {21},
        pages = {5364},
          doi = {10.3390/rs14215364},
       adsurl = {https://ui.adsabs.harvard.edu/abs/2022RemS...14.5364Z}
}

@ARTICLE{Lu_2021_O3,
       author = {{Lu}, Jinpeng and {Xie}, Fei and {Tian}, Hongying and {Luo}, Jiali},
        title = "{Impacts of Ozone Changes in the Tropopause Layer on Stratospheric Water Vapor}",
      journal = {Atmosphere},
         year = 2021,
        month = feb,
       volume = {12},
       number = {3},
          eid = {291},
        pages = {291},
          doi = {10.3390/atmos12030291},
       adsurl = {https://ui.adsabs.harvard.edu/abs/2021Atmos..12..291L}
}

@ARTICLE{2021NRvEE...2..358M,
       author = {{Mitchell}, Ross N. and {Zhang}, Nan and {Salminen}, Johanna and {Liu}, Yebo and {Spencer}, Christopher J. and {Steinberger}, Bernhard and {Murphy}, J. Brendan and {Li}, Zheng-Xiang},
        title = "{The supercontinent cycle}",
      journal = {Nature Reviews Earth and Environment},
         year = 2021,
        month = may,
       volume = {2},
       number = {5},
        pages = {358-374},
          doi = {10.1038/s43017-021-00160-0},
       adsurl = {https://ui.adsabs.harvard.edu/abs/2021NRvEE...2..358M}
}

@article{laskar2023did,
  title={Did atmospheric thermal tides cause a daylength locking in the Precambrian? A review on recent results},
  author={Laskar, Jacques and Farhat, Mohammad and Lantink, Margriet L and Auclair-Desrotour, Pierre and Bou{\'e}, Gwena{\"e}l and Sinnesael, Matthias},
  journal={arXiv preprint arXiv:2309.11479},
  year={2023}
}

@ARTICLE{2016GeoRL..43.5716B,
       author = {{Bartlett}, Benjamin C. and {Stevenson}, David J.},
        title = "{Analysis of a Precambrian resonance-stabilized day length}",
      journal = {\grl},
         year = 2016,
        month = jun,
       volume = {43},
       number = {11},
        pages = {5716-5724},
          doi = {10.1002/2016GL068912},
archivePrefix = {arXiv},
       eprint = {1502.01421},
 primaryClass = {physics.geo-ph},
       adsurl = {https://ui.adsabs.harvard.edu/abs/2016GeoRL..43.5716B}
}

@ARTICLE{1987PreR...37...95Z,
       author = {{Zahnle}, K. and {Walker}, J.~C.~G.},
        title = "{A constant daylength during the precambrian era?}",
      journal = {Precambrian Research},
         year = 1987,
        month = jan,
       volume = {37},
       number = {2},
        pages = {95-105},
          doi = {10.1016/0301-9268(87)90073-8},
       adsurl = {https://ui.adsabs.harvard.edu/abs/1987PreR...37...95Z}
}

@ARTICLE{Rogers_2002_Columbia,
       author = {{Rogers}, John J.~W. and {Santosh}, M.},
        title = "{Configuration of Columbia, a Mesoproterozoic Supercontinent}",
      journal = {Gondwana Research},
         year = 2002,
        month = jan,
       volume = {5},
       number = {1},
        pages = {5-22},
          doi = {10.1016/S1342-937X(05)70883-2},
       adsurl = {https://ui.adsabs.harvard.edu/abs/2002GondR...5....5R}
}

@ARTICLE{Forster_2007_O3,
       author = {{Forster}, Piers M. and {Bodeker}, Greg and {Schofield}, Robyn and {Solomon}, Susan and {Thompson}, David},
        title = "{Effects of ozone cooling in the tropical lower stratosphere and upper troposphere}",
      journal = {\grl},
         year = 2007,
        month = dec,
       volume = {34},
       number = {23},
          eid = {L23813},
        pages = {L23813},
          doi = {10.1029/2007GL031994},
       adsurl = {https://ui.adsabs.harvard.edu/abs/2007GeoRL..3423813F}
}

@ARTICLE{Xie_2008_O3_tape,
       author = {{Xie}, Fei and {Tian}, Wenshou and {Chipperfield}, Martyn P.},
        title = "{Radiative effect of ozone change on stratosphere-troposphere exchange}",
      journal = {Journal of Geophysical Research (Atmospheres)},
         year = 2008,
        month = apr,
       volume = {113},
       number = {D7},
          eid = {D00B09},
        pages = {D00B09},
          doi = {10.1029/2008JD009829},
       adsurl = {https://ui.adsabs.harvard.edu/abs/2008JGRD..113.0B09X}
}

@article{charette2010volume,
  title={The volume of Earth's ocean},
  author={Charette, Matthew A and Smith, Walter HF},
  journal={Oceanography},
  volume={23},
  number={2},
  pages={112--114},
  year={2010},
  publisher={JSTOR}
}

@article{kasting2025atmospheric,
  title={Atmospheric oxygen and methane on the early Earth},
  author={Kasting, James F and Ji, Aoshuang},
  journal={Philosophical Transactions B},
  volume={380},
  number={1931},
  pages={20240093},
  year={2025},
  publisher={The Royal Society}
}

@ARTICLE{Saunois_Methane_2025,
       author = {{Saunois}, Marielle and {Martinez}, Adrien and {Poulter}, Benjamin and {Zhang}, Zhen and {Raymond}, Peter A. and {Regnier}, Pierre and {Canadell}, Josep G. and {Jackson}, Robert B. and {Patra}, Prabir K. and {Bousquet}, Philippe and {Ciais}, Philippe and {Dlugokencky}, Edward J. and {Lan}, Xin and {Allen}, George H. and {Bastviken}, David and {Beerling}, David J. and {Belikov}, Dmitry A. and {Blake}, Donald R. and {Castaldi}, Simona and {Crippa}, Monica and {Deemer}, Bridget R. and {Dennison}, Fraser and {Etiope}, Giuseppe and {Gedney}, Nicola and {H{\"o}glund-Isaksson}, Lena and {Holgerson}, Meredith A. and {Hopcroft}, Peter O. and {Hugelius}, Gustaf and {Ito}, Akihiko and {Jain}, Atul K. and {Janardanan}, Rajesh and {Johnson}, Matthew S. and {Kleinen}, Thomas and {Krummel}, Paul B. and {Lauerwald}, Ronny and {Li}, Tingting and {Liu}, Xiangyu and {McDonald}, Kyle C. and {Melton}, Joe R. and {M{\"u}hle}, Jens and {M{\"u}ller}, Jurek and {Murguia-Flores}, Fabiola and {Niwa}, Yosuke and {Noce}, Sergio and {Pan}, Shufen and {Parker}, Robert J. and {Peng}, Changhui and {Ramonet}, Michel and {Riley}, William J. and {Rocher-Ros}, Gerard and {Rosentreter}, Judith A. and {Sasakawa}, Motoki and {Segers}, Arjo and {Smith}, Steven J. and {Stanley}, Emily H. and {Thanwerdas}, Jo{\"e}l and {Tian}, Hanqin and {Tsuruta}, Aki and {Tubiello}, Francesco N. and {Weber}, Thomas S. and {van der Werf}, Guido R. and {Worthy}, Douglas E.~J. and {Xi}, Yi and {Yoshida}, Yukio and {Zhang}, Wenxin and {Zheng}, Bo and {Zhu}, Qing and {Zhu}, Qiuan and {Zhuang}, Qianlai},
        title = "{Global Methane Budget 2000-2020}",
      journal = {Earth System Science Data},
         year = 2025,
        month = may,
       volume = {17},
       number = {5},
        pages = {1873-1958},
          doi = {10.5194/essd-17-1873-2025},
       adsurl = {https://ui.adsabs.harvard.edu/abs/2025ESSD...17.1873S}
}

@ARTICLE{Gu_Venus_Photo_2025,
       author = {{Gu}, Hao and {Cui}, Jun and {Wu}, Xiaoshu and {Huang}, Xu and {Wu}, Shiqi and {Li}, Wenlong and {Zhao}, Jinjin and {Lu}, Haoyu and {Li}, Lei},
        title = "{Hydrogen Loss on Venus Driven by Photochemistry}",
      journal = {\apjl},
         year = 2025,
        month = jul,
       volume = {988},
       number = {1},
          eid = {L31},
        pages = {L31},
          doi = {10.3847/2041-8213/adec90},
       adsurl = {https://ui.adsabs.harvard.edu/abs/2025ApJ...988L..31G}
}

@ARTICLE{2024ACP....24.7405Z,
       author = {{Zolghadrshojaee}, Mona and {Tegtmeier}, Susann and {Davis}, Sean M. and {Pilch Kedzierski}, Robin},
        title = "{Variability And Long-Term Changes In Tropical Cold-Point Temperature And Water Vapor}",
      journal = {Atmospheric Chemistry \& Physics},
         year = 2024,
        month = jun,
       volume = {24},
        pages = {7405-7419},
          doi = {10.5194/acp-24-7405-2024},
       adsurl = {https://ui.adsabs.harvard.edu/abs/2024ACP....24.7405Z}
}

@ARTICLE{2017JCli...30.1245L,
       author = {{Lin}, Pu and {Paynter}, David and {Ming}, Yi and {Ramaswamy}, V.},
        title = "{Changes of the Tropical Tropopause Layer under Global Warming}",
      journal = {Journal of Climate},
         year = 2017,
        month = feb,
       volume = {30},
       number = {4},
        pages = {1245-1258},
          doi = {10.1175/JCLI-D-16-0457.1},
       adsurl = {https://ui.adsabs.harvard.edu/abs/2017JCli...30.1245L}
}

@ARTICLE{2023FrEaS..1177502Z,
       author = {{Zou}, Ling and {Hoffmann}, Lars and {M{\"u}ller}, Rolf and {Spang}, Reinhold},
        title = "{Variability and trends of the tropical tropopause derived from a 1980{\textendash}2021 multi-reanalysis assessment}",
      journal = {Frontiers in Earth Science},
         year = 2023,
        month = aug,
       volume = {11},
          eid = {1177502},
        pages = {1177502},
          doi = {10.3389/feart.2023.1177502},
       adsurl = {https://ui.adsabs.harvard.edu/abs/2023FrEaS..1177502Z}
}

@ARTICLE{2023PNAS..12009751W,
       author = {{Warren}, Alexandra O. and {Kite}, Edwin S.},
        title = "{Narrow range of early habitable Venus scenarios permitted by modeling of oxygen loss and radiogenic argon degassing}",
      journal = {Proceedings of the National Academy of Science},
         year = 2023,
        month = mar,
       volume = {120},
       number = {11},
          eid = {e2209751120},
        pages = {e2209751120},
          doi = {10.1073/pnas.2209751120},
       adsurl = {https://ui.adsabs.harvard.edu/abs/2023PNAS..12009751W}
}

@ARTICLE{2021Natur.598..276T,
       author = {{Turbet}, Martin and {Bolmont}, Emeline and {Chaverot}, Guillaume and {Ehrenreich}, David and {Leconte}, J{\'e}r{\'e}my and {Marcq}, Emmanuel},
        title = "{Day-night cloud asymmetry prevents early oceans on Venus but not on Earth}",
      journal = {\nat},
         year = 2021,
        month = oct,
       volume = {598},
       number = {7880},
        pages = {276-280},
          doi = {10.1038/s41586-021-03873-w},
archivePrefix = {arXiv},
       eprint = {2110.08801},
 primaryClass = {astro-ph.EP},
       adsurl = {https://ui.adsabs.harvard.edu/abs/2021Natur.598..276T}
}

@ARTICLE{2024NatAs.tmp..289C,
       author = {{Constantinou}, Tereza and {Shorttle}, Oliver and {Rimmer}, Paul B.},
        title = "{A dry Venusian interior constrained by atmospheric chemistry}",
      journal = {Nature Astronomy},
         year = 2024,
        month = dec,
          doi = {10.1038/s41550-024-02414-5},
       adsurl = {https://ui.adsabs.harvard.edu/abs/2024NatAs.tmp..289C}
}

@article{murphy2005review,
  title={Review of the vapour pressures of ice and supercooled water for atmospheric applications},
  author={Murphy, Daniel M and Koop, Thomas},
  journal={Quarterly Journal of the Royal Meteorological Society: A journal of the atmospheric sciences, applied meteorology and physical oceanography},
  volume={131},
  number={608},
  pages={1539--1565},
  year={2005},
  publisher={Wiley Online Library}
}

@ARTICLE{2024PNAS..12119228L,
       author = {{Lin}, Jonathan and {Emanuel}, Kerry},
        title = "{Why the lower stratosphere cools when the troposphere warms}",
      journal = {Proceedings of the National Academy of Science},
         year = 2024,
        month = mar,
       volume = {121},
       number = {11},
          eid = {e2319228121},
        pages = {e2319228121},
          doi = {10.1073/pnas.2319228121},
       adsurl = {https://ui.adsabs.harvard.edu/abs/2024PNAS..12119228L}
}

@ARTICLE{2015P&SS..111...62G,
       author = {{Godolt}, M. and {Grenfell}, J.~L. and {Hamann-Reinus}, A. and {Kitzmann}, D. and {Kunze}, M. and {Langematz}, U. and {von Paris}, P. and {Patzer}, A.~B.~C. and {Rauer}, H. and {Stracke}, B.},
        title = "{3D climate modeling of Earth-like extrasolar planets orbiting different types of host stars}",
      journal = {\planss},
         year = 2015,
        month = jun,
       volume = {111},
        pages = {62-76},
          doi = {10.1016/j.pss.2015.03.010},
archivePrefix = {arXiv},
       eprint = {1504.01558},
 primaryClass = {astro-ph.EP},
       adsurl = {https://ui.adsabs.harvard.edu/abs/2015P&SS..111...62G}
}

@ARTICLE{2024MNRAS.531.1471D,
       author = {{De Luca}, P. and {Braam}, M. and {Komacek}, T.~D. and {Hochman}, A.},
        title = "{The impact of ozone on Earth-like exoplanet climate dynamics: the case of Proxima Centauri b}",
      journal = {\mnras},
         year = 2024,
        month = jun,
       volume = {531},
       number = {1},
        pages = {1471-1482},
          doi = {10.1093/mnras/stae1199},
archivePrefix = {arXiv},
       eprint = {2404.17972},
 primaryClass = {astro-ph.EP},
       adsurl = {https://ui.adsabs.harvard.edu/abs/2024MNRAS.531.1471D}
}

@ARTICLE{2025PSJ.....6....5B,
       author = {{Braam}, Marrick and {Palmer}, Paul I. and {Decin}, Leen and {Mayne}, Nathan J. and {Manners}, James and {Rugheimer}, Sarah},
        title = "{Earth-like Exoplanets in Spin{\textendash}Orbit Resonances: Climate Dynamics, 3D Atmospheric Chemistry, and Observational Signatures}",
      journal = {\psj},
         year = 2025,
        month = jan,
       volume = {6},
       number = {1},
          eid = {5},
        pages = {5},
          doi = {10.3847/PSJ/ad9565},
archivePrefix = {arXiv},
       eprint = {2410.19108},
 primaryClass = {astro-ph.EP},
       adsurl = {https://ui.adsabs.harvard.edu/abs/2025PSJ.....6....5B}
}

@ARTICLE{2019AsBio..19...99D,
       author = {{Del Genio}, Anthony D. and {Way}, Michael J. and {Amundsen}, David S. and {Aleinov}, Igor and {Kelley}, Maxwell and {Kiang}, Nancy Y. and {Clune}, Thomas L.},
        title = "{Habitable Climate Scenarios for Proxima Centauri b with a Dynamic Ocean}",
      journal = {Astrobiology},
         year = 2019,
        month = jan,
       volume = {19},
       number = {1},
        pages = {99-125},
          doi = {10.1089/ast.2017.1760},
       adsurl = {https://ui.adsabs.harvard.edu/abs/2019AsBio..19...99D}
}

@ARTICLE{2016A&A...596A.112T,
       author = {{Turbet}, Martin and {Leconte}, J{\'e}r{\'e}my and {Selsis}, Franck and {Bolmont}, Emeline and {Forget}, Fran{\c{c}}ois and {Ribas}, Ignasi and {Raymond}, Sean N. and {Anglada-Escud{\'e}}, Guillem},
        title = "{The habitability of Proxima Centauri b. II. Possible climates and observability}",
      journal = {\aap},
         year = 2016,
        month = dec,
       volume = {596},
          eid = {A112},
        pages = {A112},
          doi = {10.1051/0004-6361/201629577},
archivePrefix = {arXiv},
       eprint = {1608.06827},
 primaryClass = {astro-ph.EP},
       adsurl = {https://ui.adsabs.harvard.edu/abs/2016A&A...596A.112T}
}

@ARTICLE{2005Sci...308.1014T,
       author = {{Tian}, Feng and {Toon}, Owen B. and {Pavlov}, Alexander A. and {De Sterck}, H.},
        title = "{A Hydrogen-Rich Early Earth Atmosphere}",
      journal = {Science},
         year = 2005,
        month = may,
       volume = {308},
       number = {5724},
        pages = {1014-1017},
          doi = {10.1126/science.1106983},
       adsurl = {https://ui.adsabs.harvard.edu/abs/2005Sci...308.1014T}
}

@ARTICLE{2016ApJ...816...34O,
       author = {{Owen}, James E. and {Alvarez}, Marcelo A.},
        title = "{UV Driven Evaporation of Close-in Planets: Energy-limited, Recombination-limited, and Photon-limited Flows}",
      journal = {\apj},
         year = 2016,
        month = jan,
       volume = {816},
       number = {1},
          eid = {34},
        pages = {34},
          doi = {10.3847/0004-637X/816/1/34},
archivePrefix = {arXiv},
       eprint = {1504.07170},
 primaryClass = {astro-ph.EP},
       adsurl = {https://ui.adsabs.harvard.edu/abs/2016ApJ...816...34O}
}

@ARTICLE{2016A&A...585L...2S,
       author = {{Salz}, M. and {Schneider}, P.~C. and {Czesla}, S. and {Schmitt}, J.~H.~M.~M.},
        title = "{Energy-limited escape revised. The transition from strong planetary winds to stable thermospheres}",
      journal = {\aap},
         year = 2016,
        month = jan,
       volume = {585},
          eid = {L2},
        pages = {L2},
          doi = {10.1051/0004-6361/201527042},
archivePrefix = {arXiv},
       eprint = {1511.09348},
 primaryClass = {astro-ph.EP},
       adsurl = {https://ui.adsabs.harvard.edu/abs/2016A&A...585L...2S}
}

@ARTICLE{2017ApJ...845..130L,
       author = {{Lehmer}, Owen R. and {Catling}, David C.},
        title = "{Rocky Worlds Limited to {\ensuremath{\sim}}1.8 Earth Radii by Atmospheric Escape during a Star{\textquoteright}s Extreme UV Saturation}",
      journal = {\apj},
         year = 2017,
        month = aug,
       volume = {845},
       number = {2},
          eid = {130},
        pages = {130},
          doi = {10.3847/1538-4357/aa8137},
archivePrefix = {arXiv},
       eprint = {1706.02050},
 primaryClass = {astro-ph.EP},
       adsurl = {https://ui.adsabs.harvard.edu/abs/2017ApJ...845..130L}
}

@ARTICLE{2020GeoRL..4789533T,
       author = {{Tegtmeier}, Susann and {Anstey}, James and {Davis}, Sean and {Ivanciu}, Ioana and {Jia}, Yue and {McPhee}, David and {Pilch Kedzierski}, Robin},
        title = "{Zonal Asymmetry of the QBO Temperature Signal in the Tropical Tropopause Region}",
      journal = {\grl},
         year = 2020,
        month = dec,
       volume = {47},
       number = {24},
          eid = {e89533},
        pages = {e89533},
          doi = {10.1029/2020GL089533},
       adsurl = {https://ui.adsabs.harvard.edu/abs/2020GeoRL..4789533T}
}

@ARTICLE{2013JGRD..118.9658G,
       author = {{Garfinkel}, C.~I. and {Waugh}, D.~W. and {Oman}, L.~D. and {Wang}, L. and {Hurwitz}, M.~M.},
        title = "{Temperature trends in the tropical upper troposphere and lower stratosphere: Connections with sea surface temperatures and implications for water vapor and ozone}",
      journal = {Journal of Geophysical Research (Atmospheres)},
         year = 2013,
        month = sep,
       volume = {118},
       number = {17},
        pages = {9658-9672},
          doi = {10.1002/jgrd.50772},
       adsurl = {https://ui.adsabs.harvard.edu/abs/2013JGRD..118.9658G}
}

@ARTICLE{2012ACP....1212183P,
       author = {{Paulik}, L.~C. and {Birner}, T.},
        title = "{Quantifying the deep convective temperature signal within the tropical tropopause layer (TTL)}",
      journal = {Atmospheric Chemistry \& Physics},
         year = 2012,
        month = dec,
       volume = {12},
       number = {24},
        pages = {12183-12195},
          doi = {10.5194/acp-12-12183-201210.5194/acpd-12-19617-2012},
       adsurl = {https://ui.adsabs.harvard.edu/abs/2012ACP....1212183P}
}

@ARTICLE{2014ApJ...785L..20W,
       author = {{Wordsworth}, Robin and {Pierrehumbert}, Raymond},
        title = "{Abiotic Oxygen-dominated Atmospheres on Terrestrial Habitable Zone Planets}",
      journal = {\apjl},
         year = 2014,
        month = apr,
       volume = {785},
       number = {2},
          eid = {L20},
        pages = {L20},
          doi = {10.1088/2041-8205/785/2/L20},
archivePrefix = {arXiv},
       eprint = {1403.2713},
 primaryClass = {astro-ph.EP},
       adsurl = {https://ui.adsabs.harvard.edu/abs/2014ApJ...785L..20W}
}

@ARTICLE{2010NatGe...3..459D,
       author = {{di Achille}, Gaetano and {Hynek}, Brian M.},
        title = "{Ancient ocean on Mars supported by global distribution of deltas and valleys}",
      journal = {Nature Geoscience},
         year = 2010,
        month = jul,
       volume = {3},
       number = {7},
        pages = {459-463},
          doi = {10.1038/ngeo891},
       adsurl = {https://ui.adsabs.harvard.edu/abs/2010NatGe...3..459D}
}

@ARTICLE{2022NatAs...6.1263S,
       author = {{Sauterey}, Boris and {Charnay}, Benjamin and {Affholder}, Antonin and {Mazevet}, St{\'e}phane and {Ferri{\`e}re}, R{\'e}gis},
        title = "{Early Mars habitability and global cooling by H$_{2}$-based methanogens}",
      journal = {Nature Astronomy},
         year = 2022,
        month = nov,
       volume = {6},
        pages = {1263-1271},
          doi = {10.1038/s41550-022-01786-w},
archivePrefix = {arXiv},
       eprint = {2210.04948},
 primaryClass = {astro-ph.EP},
       adsurl = {https://ui.adsabs.harvard.edu/abs/2022NatAs...6.1263S}
}

@article{liu2025evolution,
  title={Evolution of the iodine cycle and the late stabilization of the Earth’s ozone layer},
  author={Liu, Jingjun and Hardisty, Dalton S and Kasting, James F and Fakhraee, Mojtaba and Planavsky, Noah J},
  journal={Proceedings of the National Academy of Sciences},
  volume={122},
  number={2},
  pages={e2412898121},
  year={2025},
  publisher={National Academy of Sciences}
}

@ARTICLE{2019A&G....60a1.25M,
       author = {{Mendillo}, Michael},
        title = "{The ionospheres of planets and exoplanets}",
      journal = {Astronomy and Geophysics},
         year = 2019,
        month = feb,
       volume = {60},
       number = {1},
        pages = {1.25-1.30},
          doi = {10.1093/astrogeo/atz047},
       adsurl = {https://ui.adsabs.harvard.edu/abs/2019A&G....60a1.25M}
}

@ARTICLE{2021AGUA....200323D,
       author = {{Dong}, Junjie and {Fischer}, Rebecca A. and {Stixrude}, Lars P. and {Lithgow-Bertelloni}, Carolina R.},
        title = "{Constraining the Volume of Earth's Early Oceans With a Temperature Dependent Mantle Water Storage Capacity Model}",
      journal = {AGU Advances},
         year = 2021,
        month = mar,
       volume = {2},
       number = {1},
          eid = {e2020AV000323},
        pages = {e2020AV000323},
          doi = {10.1029/2020AV000323},
       adsurl = {https://ui.adsabs.harvard.edu/abs/2021AGUA....200323D}
}

@ARTICLE{2020JGRA..12527639G,
       author = {{Gronoff}, G. and {Arras}, P. and {Baraka}, S. and {Bell}, J.~M. and {Cessateur}, G. and {Cohen}, O. and {Curry}, S.~M. and {Drake}, J.~J. and {Elrod}, M. and {Erwin}, J. and {Garcia-Sage}, K. and {Garraffo}, C. and {Glocer}, A. and {Heavens}, N.~G. and {Lovato}, K. and {Maggiolo}, R. and {Parkinson}, C.~D. and {Simon Wedlund}, C. and {Weimer}, D.~R. and {Moore}, W.~B.},
        title = "{Atmospheric Escape Processes and Planetary Atmospheric Evolution}",
      journal = {Journal of Geophysical Research (Space Physics)},
         year = 2020,
        month = aug,
       volume = {125},
       number = {8},
          eid = {e27639},
        pages = {e27639},
          doi = {10.1029/2019JA02763910.1002/essoar.10502458.1},
       adsurl = {https://ui.adsabs.harvard.edu/abs/2020JGRA..12527639G}
}

@ARTICLE{2012PNAS..109.4371P,
       author = {{Pope}, Emily C. and {Bird}, Dennis K. and {Rosing}, Minik T.},
        title = "{Isotope composition and volume of Earth's early oceans}",
      journal = {Proceedings of the National Academy of Science},
         year = 2012,
        month = mar,
       volume = {109},
       number = {12},
        pages = {4371-4376},
          doi = {10.1073/pnas.1115705109},
       adsurl = {https://ui.adsabs.harvard.edu/abs/2012PNAS..109.4371P}
}

@ARTICLE{2018E&PSL.497..149K,
       author = {{Kurokawa}, Hiroyuki and {Foriel}, Julien and {Laneuville}, Matthieu and {Houser}, Christine and {Usui}, Tomohiro},
        title = "{Subduction and atmospheric escape of Earth's seawater constrained by hydrogen isotopes}",
      journal = {Earth and Planetary Science Letters},
         year = 2018,
        month = sep,
       volume = {497},
        pages = {149-160},
          doi = {10.1016/j.epsl.2018.06.016},
archivePrefix = {arXiv},
       eprint = {1806.03792},
 primaryClass = {astro-ph.EP},
       adsurl = {https://ui.adsabs.harvard.edu/abs/2018E&PSL.497..149K}
}

@ARTICLE{2017RSPTA.37550393K,
       author = {{Korenaga}, Jun and {Planavsky}, Noah J. and {Evans}, David A.~D.},
        title = "{Global water cycle and the coevolution of the Earth's interior and surface environment}",
      journal = {Philosophical Transactions of the Royal Society of London Series A},
         year = 2017,
        month = apr,
       volume = {375},
       number = {2094},
          eid = {20150393},
        pages = {20150393},
          doi = {10.1098/rsta.2015.0393},
       adsurl = {https://ui.adsabs.harvard.edu/abs/2017RSPTA.37550393K}
}

@ARTICLE{2003ARA&A..41..429K,
       author = {{Kasting}, James F. and {Catling}, David},
        title = "{Evolution of a Habitable Planet}",
      journal = {\araa},
         year = 2003,
        month = jan,
       volume = {41},
        pages = {429-463},
          doi = {10.1146/annurev.astro.41.071601.170049},
       adsurl = {https://ui.adsabs.harvard.edu/abs/2003ARA&A..41..429K}
}

@ARTICLE{2018Sci...361..490O,
       author = {{Orosei}, R. and {Lauro}, S.~E. and {Pettinelli}, E. and {Cicchetti}, A. and {Coradini}, M. and {Cosciotti}, B. and {Di Paolo}, F. and {Flamini}, E. and {Mattei}, E. and {Pajola}, M. and {Soldovieri}, F. and {Cartacci}, M. and {Cassenti}, F. and {Frigeri}, A. and {Giuppi}, S. and {Martufi}, R. and {Masdea}, A. and {Mitri}, G. and {Nenna}, C. and {Noschese}, R. and {Restano}, M. and {Seu}, R.},
        title = "{Radar evidence of subglacial liquid water on Mars}",
      journal = {Science},
         year = 2018,
        month = aug,
       volume = {361},
       number = {6401},
        pages = {490-493},
          doi = {10.1126/science.aar7268},
archivePrefix = {arXiv},
       eprint = {2004.04587},
 primaryClass = {astro-ph.EP},
       adsurl = {https://ui.adsabs.harvard.edu/abs/2018Sci...361..490O}
}

@ARTICLE{2021NatAs...5...63L,
       author = {{Lauro}, Sebastian Emanuel and {Pettinelli}, Elena and {Caprarelli}, Graziella and {Guallini}, Luca and {Rossi}, Angelo Pio and {Mattei}, Elisabetta and {Cosciotti}, Barbara and {Cicchetti}, Andrea and {Soldovieri}, Francesco and {Cartacci}, Marco and {Di Paolo}, Federico and {Noschese}, Raffaella and {Orosei}, Roberto},
        title = "{Multiple subglacial water bodies below the south pole of Mars unveiled by new MARSIS data}",
      journal = {Nature Astronomy},
         year = 2021,
        month = jan,
       volume = {5},
        pages = {63-70},
          doi = {10.1038/s41550-020-1200-6},
archivePrefix = {arXiv},
       eprint = {2010.00870},
 primaryClass = {astro-ph.EP},
       adsurl = {https://ui.adsabs.harvard.edu/abs/2021NatAs...5...63L}
}

@ARTICLE{2007SSRv..129..245L,
       author = {{Lundin}, Rickard and {Lammer}, Helmut and {Ribas}, Ignasi},
        title = "{Planetary Magnetic Fields and Solar Forcing: Implications for Atmospheric Evolution}",
      journal = {\ssr},
         year = 2007,
        month = mar,
       volume = {129},
       number = {1-3},
        pages = {245-278},
          doi = {10.1007/s11214-007-9176-4},
       adsurl = {https://ui.adsabs.harvard.edu/abs/2007SSRv..129..245L}
}

@ARTICLE{2018NatAs...2..287M,
       author = {{Mendillo}, Michael and {Withers}, Paul and {Dalba}, Paul A.},
        title = "{Atomic oxygen ions as ionospheric biomarkers on exoplanets}",
      journal = {Nature Astronomy},
         year = 2018,
        month = feb,
       volume = {2},
        pages = {287-291},
          doi = {10.1038/s41550-017-0375-y},
       adsurl = {https://ui.adsabs.harvard.edu/abs/2018NatAs...2..287M}
}

@ARTICLE{1973Icar...20..437D,
       author = {{Danielson}, R.~E. and {Caldwell}, John J. and {Larach}, D.~R.},
        title = "{An Inversion in the Atmosphere of Titan}",
      journal = {\icarus},
         year = 1973,
        month = dec,
       volume = {20},
       number = {4},
        pages = {437-443},
          doi = {10.1016/0019-1035(73)90016-X},
       adsurl = {https://ui.adsabs.harvard.edu/abs/1973Icar...20..437D}
}

@BOOK{2017aeil.book.....C,
       author = {{Catling}, David C. and {Kasting}, James F.},
        title = "{Atmospheric Evolution on Inhabited and Lifeless Worlds}",
         year = 2017,
       publisher = {Cambridge University Press},
       adsurl = {https://ui.adsabs.harvard.edu/abs/2017aeil.book.....C}
}

@ARTICLE{2010ApJ...716.1573J,
       author = {{Johnson}, Robert E.},
        title = "{Thermally Driven Atmospheric Escape}",
      journal = {\apj},
         year = 2010,
        month = jun,
       volume = {716},
       number = {2},
        pages = {1573-1578},
          doi = {10.1088/0004-637X/716/2/1573},
archivePrefix = {arXiv},
       eprint = {1001.0917},
 primaryClass = {astro-ph.EP},
       adsurl = {https://ui.adsabs.harvard.edu/abs/2010ApJ...716.1573J}
}

@ARTICLE{2021NatGe..14..138O,
       author = {{Ozaki}, Kazumi and {Reinhard}, Christopher T.},
        title = "{The future lifespan of Earth's oxygenated atmosphere}",
      journal = {Nature Geoscience},
         year = 2021,
        month = jan,
       volume = {14},
       number = {3},
        pages = {138-142},
          doi = {10.1038/s41561-021-00693-5},
archivePrefix = {arXiv},
       eprint = {2103.02694},
 primaryClass = {astro-ph.EP},
       adsurl = {https://ui.adsabs.harvard.edu/abs/2021NatGe..14..138O}
}

@article{wright2024liquid,
  title={Liquid water in the Martian mid-crust},
  author={Wright, Vashan and Morzfeld, Matthias and Manga, Michael},
  journal={Proceedings of the National Academy of Sciences},
  volume={121},
  number={35},
  pages={e2409983121},
  year={2024},
  publisher={National Academy of Sciences}
}

@article{krause2022extreme,
  title={Extreme variability in atmospheric oxygen levels in the late Precambrian},
  author={Krause, Alexander J and Mills, Benjamin JW and Merdith, Andrew S and Lenton, Timothy M and Poulton, Simon W},
  journal={Science advances},
  volume={8},
  number={41},
  pages={eabm8191},
  year={2022},
  publisher={American Association for the Advancement of Science}
}

@article{yierpan2020recycled,
  title={Recycled selenium in hot spot--influenced lavas records ocean-atmosphere oxygenation},
  author={Yierpan, Aierken and K{\"o}nig, Stephan and Labidi, Jabrane and Schoenberg, Ronny},
  journal={Science Advances},
  volume={6},
  number={39},
  pages={eabb6179},
  year={2020},
  publisher={American Association for the Advancement of Science}
}

@ARTICLE{2018GeCoA.232...82A,
       author = {{Avice}, G. and {Marty}, B. and {Burgess}, R. and {Hofmann}, A. and {Philippot}, P. and {Zahnle}, K. and {Zakharov}, D.},
        title = "{Evolution of atmospheric xenon and other noble gases inferred from Archean to Paleoproterozoic rocks}",
      journal = {\gca},
         year = 2018,
        month = jul,
       volume = {232},
        pages = {82-100},
          doi = {10.1016/j.gca.2018.04.018},
       adsurl = {https://ui.adsabs.harvard.edu/abs/2018GeCoA.232...82A}
}

@ARTICLE{2018PhyU...61..217S,
       author = {{Shematovich}, V.~I. and {Marov}, M. Ya},
        title = "{Escape of planetary atmospheres: physical processes and numerical models}",
      journal = {Physics Uspekhi},
         year = 2018,
        month = mar,
       volume = {61},
       number = {3},
        pages = {217},
          doi = {10.3367/UFNe.2017.09.038212},
       adsurl = {https://ui.adsabs.harvard.edu/abs/2018PhyU...61..217S}
}

@ARTICLE{2021NatGe..14..143G,
       author = {{Goldblatt}, Colin and {McDonald}, Victoria L. and {McCusker}, Kelly E.},
        title = "{Earth's long-term climate stabilized by clouds}",
      journal = {Nature Geoscience},
         year = 2021,
        month = jan,
       volume = {14},
       number = {3},
        pages = {143-150},
          doi = {10.1038/s41561-021-00691-7},
       adsurl = {https://ui.adsabs.harvard.edu/abs/2021NatGe..14..143G}
}

@ARTICLE{2020SSRv..216...90C,
       author = {{Charnay}, Benjamin and {Wolf}, Eric T. and {Marty}, Bernard and {Forget}, Fran{\c{c}}ois},
        title = "{Is the Faint Young Sun Problem for Earth Solved?}",
      journal = {\ssr},
         year = 2020,
        month = jul,
       volume = {216},
       number = {5},
          eid = {90},
        pages = {90},
          doi = {10.1007/s11214-020-00711-9},
archivePrefix = {arXiv},
       eprint = {2006.06265},
 primaryClass = {astro-ph.EP},
       adsurl = {https://ui.adsabs.harvard.edu/abs/2020SSRv..216...90C}
}

@ARTICLE{2019PNAS..11617207H,
       author = {{Hodgskiss}, Malcolm S.~W. and {Crockford}, Peter W. and {Peng}, Yongbo and {Wing}, Boswell A. and {Horner}, Tristan J.},
        title = "{A productivity collapse to end Earth's Great Oxidation}",
      journal = {Proceedings of the National Academy of Science},
         year = 2019,
        month = aug,
       volume = {116},
       number = {35},
        pages = {17207-17212},
          doi = {10.1073/pnas.1900325116},
       adsurl = {https://ui.adsabs.harvard.edu/abs/2019PNAS..11617207H}
}

@ARTICLE{2020PNAS..11713314W,
       author = {{Warke}, Matthew R. and {Di Rocco}, Tommaso and {Zerkle}, Aubrey L. and {Lepland}, Aivo and {Prave}, Anthony R. and {Martin}, Adam P. and {Ueno}, Yuichiro and {Condon}, Daniel J. and {Claire}, Mark W.},
        title = "{The Great Oxidation Event preceded a Paleoproterozoic ``snowball Earth''}",
      journal = {Proceedings of the National Academy of Science},
         year = 2020,
        month = jun,
       volume = {117},
       number = {24},
        pages = {13314-13320},
          doi = {10.1073/pnas.2003090117},
       adsurl = {https://ui.adsabs.harvard.edu/abs/2020PNAS..11713314W}
}

@ARTICLE{2002GeCoA..66.3811H,
       author = {{Holland}, Heinrich D.},
        title = "{Volcanic gases, black smokers, and the great oxidation event}",
      journal = {\gca},
         year = 2002,
        month = nov,
       volume = {66},
       number = {21},
        pages = {3811-3826},
          doi = {10.1016/S0016-7037(02)00950-X},
       adsurl = {https://ui.adsabs.harvard.edu/abs/2002GeCoA..66.3811H}
}

@ARTICLE{2012JGRD..117.4211W,
       author = {{Wang}, Tao and {Dessler}, Andrew E.},
        title = "{Analysis of cirrus in the tropical tropopause layer from CALIPSO and MLS data: A water perspective}",
      journal = {Journal of Geophysical Research (Atmospheres)},
         year = 2012,
        month = feb,
       volume = {117},
       number = {D4},
          eid = {D04211},
        pages = {D04211},
          doi = {10.1029/2011JD016442},
       adsurl = {https://ui.adsabs.harvard.edu/abs/2012JGRD..117.4211W}
}

@ARTICLE{2016E&PSL.434...42D,
       author = {{Daines}, Stuart J. and {Lenton}, Timothy M.},
        title = "{The effect of widespread early aerobic marine ecosystems on methane cycling and the Great Oxidation}",
      journal = {Earth and Planetary Science Letters},
         year = 2016,
        month = jan,
       volume = {434},
        pages = {42-51},
          doi = {10.1016/j.epsl.2015.11.021},
       adsurl = {https://ui.adsabs.harvard.edu/abs/2016E&PSL.434...42D}
}

@ARTICLE{2018Geo....46..139Z,
       author = {{Zhao}, Mingyu and {Reinhard}, Christopher T. and {Planavsky}, Noah},
        title = "{Terrestrial methane fluxes and Proterozoic climate}",
      journal = {Geology},
         year = 2018,
        month = feb,
       volume = {46},
       number = {2},
        pages = {139-142},
          doi = {10.1130/G39502.1},
       adsurl = {https://ui.adsabs.harvard.edu/abs/2018Geo....46..139Z}
}

@ARTICLE{2020SciA....6.1420C,
       author = {{Catling}, David C. and {Zahnle}, Kevin J.},
        title = "{The Archean atmosphere}",
      journal = {Science Advances},
         year = 2020,
        month = feb,
       volume = {6},
       number = {9},
          eid = {eaax1420},
        pages = {eaax1420},
          doi = {10.1126/sciadv.aax1420},
       adsurl = {https://ui.adsabs.harvard.edu/abs/2020SciA....6.1420C}
}

@ARTICLE{2014Natur.506..307L,
       author = {{Lyons}, Timothy W. and {Reinhard}, Christopher T. and {Planavsky}, Noah J.},
        title = "{The rise of oxygen in Earth's early ocean and atmosphere}",
      journal = {\nat},
         year = 2014,
        month = feb,
       volume = {506},
       number = {7488},
        pages = {307-315},
          doi = {10.1038/nature13068},
       adsurl = {https://ui.adsabs.harvard.edu/abs/2014Natur.506..307L}
}

@ARTICLE{2020PreR..34305722S,
       author = {{Steadman}, J.~A. and {Large}, R.~R. and {Blamey}, N.~J. and {Mukherjee}, I. and {Corkrey}, R. and {Danyushevsky}, L.~V. and {Maslennikov}, V. and {Hollings}, P. and {Garven}, G. and {Brand}, U. and {L{\'e}cuyer}, C.},
        title = "{Evidence for elevated and variable atmospheric oxygen in the Precambrian}",
      journal = {Precambrian Research},
         year = 2020,
        month = jul,
       volume = {343},
        pages = {105722},
          doi = {10.1016/j.precamres.2020.105722},
       adsurl = {https://ui.adsabs.harvard.edu/abs/2020PreR..34305722S}
}

@ARTICLE{2019E&PSL.522...48L,
       author = {{Laakso}, Thomas A. and {Schrag}, Daniel P.},
        title = "{Methane in the Precambrian atmosphere}",
      journal = {Earth and Planetary Science Letters},
         year = 2019,
        month = sep,
       volume = {522},
        pages = {48-54},
          doi = {10.1016/j.epsl.2019.06.022},
       adsurl = {https://ui.adsabs.harvard.edu/abs/2019E&PSL.522...48L}
}

@ARTICLE{2003Geo....31...87P,
       author = {{Pavlov}, Alexander A. and {Hurtgen}, Matthew T. and {Kasting}, James F. and {Arthur}, Michael A.},
        title = "{Methane-rich Proterozoic atmosphere?}",
      journal = {Geology},
         year = 2003,
        month = jan,
       volume = {31},
       number = {1},
        pages = {87},
          doi = {10.1130/0091-7613(2003)031<0087:MRPA>2.0.CO;2},
       adsurl = {https://ui.adsabs.harvard.edu/abs/2003Geo....31...87P}
}

@ARTICLE{2022WCD.....3..139I,
       author = {{Ivanciu}, Ioana and {Matthes}, Katja and {Biastoch}, Arne and {Wahl}, Sebastian and {Harla{\ss}}, Jan},
        title = "{Twenty-first-century Southern Hemisphere impacts of ozone recovery and climate change from the stratosphere to the ocean}",
      journal = {Weather and Climate Dynamics},
         year = 2022,
        month = feb,
       volume = {3},
       number = {1},
        pages = {139-171},
          doi = {10.5194/wcd-3-139-2022},
       adsurl = {https://ui.adsabs.harvard.edu/abs/2022WCD.....3..139I}
}

@misc{fahey2018scientific,
  title={Scientific assessment of ozone depletion: 2018, global ozone research and monitoring project-report no. 58},
  author={Fahey, David and Newman, Paul A and Pyle, John A and Safari, Bonfils and Chipperfield, Martyn P and Karoly, David and Kinnison, Doug E and Ko, Malcolm and Santee, Michelle and Doherty, Sarah J},
  year={2018},
  publisher={World Meteorological Organization}
}

@ARTICLE{2002JGRD..107.4049R,
       author = {{Rosenfield}, Joan E. and {Douglass}, Anne R. and {Considine}, David B.},
        title = "{The impact of increasing carbon dioxide on ozone recovery}",
      journal = {Journal of Geophysical Research (Atmospheres)},
         year = 2002,
        month = mar,
       volume = {107},
       number = {D6},
          eid = {4049},
        pages = {4049},
          doi = {10.1029/2001JD000824},
       adsurl = {https://ui.adsabs.harvard.edu/abs/2002JGRD..107.4049R}
}

@ARTICLE{2021ACP....2111041M,
       author = {{Maliniemi}, Ville and {Nesse Tyss{\o}y}, Hilde and {Smith-Johnsen}, Christine and {Arsenovic}, Pavle and {Marsh}, Daniel R.},
        title = "{Effects of enhanced downwelling of NO$_{x}$ on Antarctic upper-stratospheric ozone in the 21st century}",
      journal = {Atmospheric Chemistry \& Physics},
         year = 2021,
        month = jul,
       volume = {21},
       number = {14},
        pages = {11041-11052},
          doi = {10.5194/acp-21-11041-2021},
       adsurl = {https://ui.adsabs.harvard.edu/abs/2021ACP....2111041M}
}

@misc{banks1973aeronomy,
  title={Aeronomy, volume 2, 263 pp},
  author={Banks, PM and Kockarts, G},
  year={1973},
  publisher={Academic, San Diego, Calif}
}

@ARTICLE{ji2024correlated,
       author = {{Ji}, Aoshuang and {Tomazzeli}, Orlando G. and {Palancar}, Gustavo G. and {Chaverot}, Guillaume and {Barker}, Mackenzie and {Fern{\'a}ndez}, Rafael P. and {Minschwaner}, Kenneth and {Kasting}, James F.},
        title = "{A Correlated-K Parameterization for O$_{2}$ Photolysis in the Schumann-Runge Bands}",
      journal = {Journal of Geophysical Research (Atmospheres)},
         year = 2024,
        month = may,
       volume = {129},
       number = {10},
          eid = {e2023JD040610},
        pages = {e2023JD040610},
          doi = {10.1029/2023JD040610},
       adsurl = {https://ui.adsabs.harvard.edu/abs/2024JGRD..12940610J}
}

@ARTICLE{2019JGRD..12411819G,
       author = {{Graham}, R.~J. and {Shaw}, Tiffany A. and {Abbot}, Dorian S.},
        title = "{The Snowball Stratosphere}",
      journal = {Journal of Geophysical Research (Atmospheres)},
         year = 2019,
        month = nov,
       volume = {124},
       number = {22},
        pages = {11,819-11,836},
          doi = {10.1029/2019JD031361},
archivePrefix = {arXiv},
       eprint = {1909.12717},
 primaryClass = {physics.ao-ph},
       adsurl = {https://ui.adsabs.harvard.edu/abs/2019JGRD..12411819G}
}

@BOOK{1977evat.book.....W,
       author = {{Walker}, James C.~G.},
        title = "{Evolution of the atmosphere}",
         year = 1977,
       adsurl = {https://ui.adsabs.harvard.edu/abs/1977evat.book.....W},
      publisher={Macmillan}
}

@ARTICLE{2022RSOS....911165C,
       author = {{Cooke}, G.~J. and {Marsh}, D.~R. and {Walsh}, C. and {Black}, B. and {Lamarque}, J. -F.},
        title = "{A revised lower estimate of ozone columns during Earth's oxygenated history}",
      journal = {Royal Society Open Science},
         year = 2022,
        month = jan,
       volume = {9},
       number = {1},
          eid = {211165},
        pages = {211165},
          doi = {10.1098/rsos.211165},
archivePrefix = {arXiv},
       eprint = {2102.11675},
 primaryClass = {astro-ph.EP},
       adsurl = {https://ui.adsabs.harvard.edu/abs/2022RSOS....911165C}
}

@INCOLLECTION{2018haex.bookE.189O,
       author = {{Olson}, Stephanie L. and {Schwieterman}, Edward W. and {Reinhard}, Christopher T. and {Lyons}, Timothy W.},
        title = "{Earth: Atmospheric Evolution of a Habitable Planet}",
    booktitle = {Handbook of Exoplanets},
         year = 2018,
         publisher = {Springer},
       editor = {{Deeg}, Hans J. and {Belmonte}, Juan Antonio},
          eid = {189},
        pages = {189},
          doi = {10.1007/978-3-319-55333-7_189},
       adsurl = {https://ui.adsabs.harvard.edu/abs/2018haex.bookE.189O}
}

@article{dunn2013evolution,
  title={Evolution: out of the ocean},
  author={Dunn, Casey W},
  journal={Current Biology},
  volume={23},
  number={6},
  pages={R241--R243},
  year={2013},
  publisher={Elsevier}
}

@ARTICLE{2021AGUA....200294K,
       author = {{Krissansen-Totton}, Joshua and {Fortney}, Jonathan J. and {Nimmo}, Francis and {Wogan}, Nicholas},
        title = "{Oxygen False Positives on Habitable Zone Planets Around Sun-Like Stars}",
      journal = {AGU Advances},
         year = 2021,
        month = jun,
       volume = {2},
       number = {2},
          eid = {e00294},
        pages = {e00294},
          doi = {10.1029/2020AV000294},
archivePrefix = {arXiv},
       eprint = {2104.06463},
 primaryClass = {astro-ph.EP},
       adsurl = {https://ui.adsabs.harvard.edu/abs/2021AGUA....200294K}
}

@ARTICLE{2022ApJ...933..115K,
       author = {{Krissansen-Totton}, J. and {Fortney}, J.~J.},
        title = "{Predictions for Observable Atmospheres of Trappist-1 Planets from a Fully Coupled Atmosphere-Interior Evolution Model}",
      journal = {\apj},
         year = 2022,
        month = jul,
       volume = {933},
       number = {1},
          eid = {115},
        pages = {115},
          doi = {10.3847/1538-4357/ac69cb},
archivePrefix = {arXiv},
       eprint = {2207.04164},
 primaryClass = {astro-ph.EP},
       adsurl = {https://ui.adsabs.harvard.edu/abs/2022ApJ...933..115K}
}

@ARTICLE{2018AsBio..18..856G,
       author = {{Gebauer}, S. and {Grenfell}, J.~L. and {Lehmann}, R. and {Rauer}, H.},
        title = "{Evolution of Earth-like Planetary Atmospheres around M Dwarf Stars: Assessing the Atmospheres and Biospheres with a Coupled Atmosphere Biogeochemical Model}",
      journal = {Astrobiology},
         year = 2018,
        month = jul,
       volume = {18},
       number = {7},
        pages = {856-872},
          doi = {10.1089/ast.2017.1723},
archivePrefix = {arXiv},
       eprint = {1807.10616},
 primaryClass = {astro-ph.EP},
       adsurl = {https://ui.adsabs.harvard.edu/abs/2018AsBio..18..856G}
}

@ARTICLE{1973JAtS...30.1481H,
       author = {{Hunten}, Donald M.},
        title = "{The Escape of Light Gases from Planetary Atmospheres.}",
      journal = {Journal of the Atmospheric Sciences},
         year = 1973,
        month = nov,
       volume = {30},
       number = {8},
        pages = {1481-1494},
          doi = {10.1175/1520-0469(1973)030<1481:TEOLGF>2.0.CO;2},
       adsurl = {https://ui.adsabs.harvard.edu/abs/1973JAtS...30.1481H}
}

@ARTICLE{2018JAMES..10..381L,
       author = {{Liu}, Han-Li and {Bardeen}, Charles G. and {Foster}, Benjamin T. and {Lauritzen}, Peter and {Liu}, Jing and {Lu}, Gang and {Marsh}, Daniel R. and {Maute}, Astrid and {McInerney}, Joseph M. and {Pedatella}, Nicholas M. and {Qian}, Liying and {Richmond}, Arthur D. and {Roble}, Raymond G. and {Solomon}, Stanley C. and {Vitt}, Francis M. and {Wang}, Wenbin},
        title = "{Development and Validation of the Whole Atmosphere Community Climate Model With Thermosphere and Ionosphere Extension (WACCM-X 2.0)}",
      journal = {Journal of Advances in Modeling Earth Systems},
         year = 2018,
        month = feb,
       volume = {10},
       number = {2},
        pages = {381-402},
          doi = {10.1002/2017MS001232},
       adsurl = {https://ui.adsabs.harvard.edu/abs/2018JAMES..10..381L}
}

@ARTICLE{1998GeoRL..25.4165D,
       author = {{Dessler}, A.~E.},
        title = "{A reexamination of the {\textquotedblleft}stratospheric fountain{\textquotedblright} hypothesis}",
      journal = {\grl},
         year = 1998,
        month = nov,
       volume = {25},
       number = {22},
        pages = {4165-4168},
          doi = {10.1029/1998GL900120},
       adsurl = {https://ui.adsabs.harvard.edu/abs/1998GeoRL..25.4165D}
}

@ARTICLE{Graham_2024_Extension,
       author = {{Graham}, R.~J. and {Halevy}, Itay and {Abbot}, Dorian},
        title = "{Substantial Extension of the Lifetime of the Terrestrial Biosphere}",
      journal = {\psj},
         year = 2024,
        month = nov,
       volume = {5},
       number = {11},
          eid = {255},
        pages = {255},
          doi = {10.3847/PSJ/ad7856},
archivePrefix = {arXiv},
       eprint = {2409.10714},
 primaryClass = {astro-ph.EP},
       adsurl = {https://ui.adsabs.harvard.edu/abs/2024PSJ.....5..255G}
}

@ARTICLE{2024ComEE...5..609L,
       author = {{Luo}, Anbo and {Sun}, Guangyi and {Grasby}, Stephen E. and {Yin}, Runsheng},
        title = "{Large igneous provinces played a major role in oceanic oxygenation events during the mid-Proterozoic}",
      journal = {Communications Earth and Environment},
         year = 2024,
        month = oct,
       volume = {5},
       number = {1},
          eid = {609},
        pages = {609},
          doi = {10.1038/s43247-024-01780-2},
       adsurl = {https://ui.adsabs.harvard.edu/abs/2024ComEE...5..609L}
}

@ARTICLE{2023AREPS..51..253M,
       author = {{Mills}, Benjamin J.~W. and {Krause}, Alexander J. and {Jarvis}, Ian and {Cramer}, Bradley D.},
        title = "{Evolution of Atmospheric O$_{2}$ Through the Phanerozoic, Revisited}",
      journal = {Annual Review of Earth and Planetary Sciences},
         year = 2023,
        month = may,
       volume = {51},
        pages = {253-276},
          doi = {10.1146/annurev-earth-032320-095425},
       adsurl = {https://ui.adsabs.harvard.edu/abs/2023AREPS..51..253M}
}

@ARTICLE{2023GSAB..135..753X,
       author = {{Xie}, Baozeng and {Zhu}, Jian-ming and {Wang}, Xiangli and {Xu}, Dongtao and {Zhou}, Limin and {Zhou}, Xiqiang and {Shi}, Xiaoying and {Tang}, Dongjie},
        title = "{Mesoproterozoic oxygenation event: From shallow marine to atmosphere}",
      journal = {Geological Society of America Bulletin},
         year = 2023,
        month = mar,
       volume = {135},
       number = {3-4},
        pages = {753-766},
          doi = {10.1130/B36407.1},
       adsurl = {https://ui.adsabs.harvard.edu/abs/2023GSAB..135..753X}
}

@ARTICLE{2006GeoRL..3312811S,
       author = {{Schoeberl}, M.~R. and {Duncan}, B.~N. and {Douglass}, A.~R. and {Waters}, J. and {Livesey}, N. and {Read}, W. and {Filipiak}, M.},
        title = "{The carbon monoxide tape recorder}",
      journal = {\grl},
         year = 2006,
        month = jun,
       volume = {33},
       number = {12},
          eid = {L12811},
        pages = {L12811},
          doi = {10.1029/2006GL026178},
       adsurl = {https://ui.adsabs.harvard.edu/abs/2006GeoRL..3312811S}
}

@ARTICLE{2008GeoRL..35.5801P,
       author = {{Pumphrey}, H.~C. and {Boone}, C. and {Walker}, K.~A. and {Bernath}, P. and {Livesey}, N.~J.},
        title = "{Tropical tape recorder observed in HCN}",
      journal = {\grl},
         year = 2008,
        month = mar,
       volume = {35},
       number = {5},
          eid = {L05801},
        pages = {L05801},
          doi = {10.1029/2007GL032137},
       adsurl = {https://ui.adsabs.harvard.edu/abs/2008GeoRL..35.5801P}
}

@ARTICLE{2024NatGe..17..458A,
       author = {{Alcott}, Lewis J. and {Walton}, Craig and {Planavsky}, Noah J. and {Shorttle}, Oliver and {Mills}, Benjamin J.~W.},
        title = "{Crustal carbonate build-up as a driver for Earth's oxygenation}",
      journal = {Nature Geoscience},
         year = 2024,
        month = may,
       volume = {17},
       number = {5},
        pages = {458-464},
          doi = {10.1038/s41561-024-01417-1},
       adsurl = {https://ui.adsabs.harvard.edu/abs/2024NatGe..17..458A}
}

@ARTICLE{2024NatGe..17..667S,
       author = {{Stockey}, Richard G. and {Cole}, Devon B. and {Farrell}, Una C. and {Agi{\'c}}, Heda and {Boag}, Thomas H. and {Brocks}, Jochen J. and {Canfield}, Don E. and {Cheng}, Meng and {Crockford}, Peter W. and {Cui}, Huan and {Dahl}, Tais W. and {Del Mouro}, Lucas and {Dewing}, Keith and {Dornbos}, Stephen Q. and {Emmings}, Joseph F. and {Gaines}, Robert R. and {Gibson}, Timothy M. and {Gill}, Benjamin C. and {Gilleaudeau}, Geoffrey J. and {Goldberg}, Karin and {Guilbaud}, Romain and {Halverson}, Galen and {Hammarlund}, Emma U. and {Hantsoo}, Kalev and {Henderson}, Miles A. and {Henderson}, Charles M. and {Hodgskiss}, Malcolm S.~W. and {Jarrett}, Amber J.~M. and {Johnston}, David T. and {Kabanov}, Pavel and {Kimmig}, Julien and {Knoll}, Andrew H. and {Kunzmann}, Marcus and {LeRoy}, Matthew A. and {Li}, Chao and {Loydell}, David K. and {Macdonald}, Francis A. and {Magnall}, Joseph M. and {Mills}, N. Tanner and {Och}, Lawrence M. and {O'Connell}, Brennan and {Pag{\`e}s}, Anais and {Peters}, Shanan E. and {Porter}, Susannah M. and {Poulton}, Simon W. and {Ritzer}, Samantha R. and {Rooney}, Alan D. and {Schoepfer}, Shane and {Smith}, Emily F. and {Strauss}, Justin V. and {Uhlein}, Gabriel Jub{\'e} and {White}, Tristan and {Wood}, Rachel A. and {Woltz}, Christina R. and {Yurchenko}, Inessa and {Planavsky}, Noah J. and {Sperling}, Erik A.},
        title = "{Sustained increases in atmospheric oxygen and marine productivity in the Neoproterozoic and Palaeozoic eras}",
      journal = {Nature Geoscience},
         year = 2024,
        month = jul,
       volume = {17},
       number = {7},
        pages = {667-674},
          doi = {10.1038/s41561-024-01479-1},
       adsurl = {https://ui.adsabs.harvard.edu/abs/2024NatGe..17..667S}
}

@ARTICLE{2024NatCo..15.6794F,
       author = {{Fakhraee}, Mojtaba and {Planavsky}, Noah},
        title = "{Insights from a dynamical system approach into the history of atmospheric oxygenation}",
      journal = {Nature Communications},
         year = 2024,
        month = aug,
       volume = {15},
          eid = {6794},
        pages = {6794},
          doi = {10.1038/s41467-024-51042-0},
       adsurl = {https://ui.adsabs.harvard.edu/abs/2024NatCo..15.6794F}
}

@ARTICLE{2020GeoRL..4786320D,
       author = {{Dong}, W.~H. and {Lin}, Y.~L. and {Zhang}, M.~H. and {Huang}, X.~M.},
        title = "{Footprint of Tropical Mesoscale Convective System Variability on Stratospheric Water Vapor}",
      journal = {\grl},
         year = 2020,
        month = mar,
       volume = {47},
       number = {5},
          eid = {e86320},
        pages = {e86320},
          doi = {10.1029/2019GL086320},
       adsurl = {https://ui.adsabs.harvard.edu/abs/2020GeoRL..4786320D}
}

@article{newell1981stratospheric,
  title={A stratospheric fountain?},
  author={Newell, Reginald E and Gould-Stewart, Sharon},
  journal={Journal of Atmospheric Sciences},
  volume={38},
  number={12},
  pages={2789--2796},
  year={1981}
}

@ARTICLE{1988Icar...74..472K,
       author = {{Kasting}, J.~F.},
        title = "{Runaway and moist greenhouse atmospheres and the evolution of Earth and Venus}",
      journal = {\icarus},
         year = 1988,
        month = jun,
       volume = {74},
       number = {3},
        pages = {472-494},
          doi = {10.1016/0019-1035(88)90116-9},
       adsurl = {https://ui.adsabs.harvard.edu/abs/1988Icar...74..472K}
}

@ARTICLE{1982GeoRL...9..605D,
       author = {{Danielsen}, Edwin F.},
        title = "{A dehydration mechanism for the stratosphere}",
      journal = {\grl},
         year = 1982,
        month = jun,
       volume = {9},
       number = {6},
        pages = {605-608},
          doi = {10.1029/GL009i006p00605},
       adsurl = {https://ui.adsabs.harvard.edu/abs/1982GeoRL...9..605D}
}

@ARTICLE{2007ACPD....7.8933P,
       author = {{Pommereau}, J. -P. and {Held}, G.},
        title = "{Is there a stratospheric fountain?}",
      journal = {Atmospheric Chemistry \& Physics Discussions},
         year = 2007,
        month = jun,
       volume = {7},
       number = {3},
        pages = {8933-8950},
          doi = {10.5194/acpd-7-8933-2007},
       adsurl = {https://ui.adsabs.harvard.edu/abs/2007ACPD....7.8933P}
}

@ARTICLE{2021ApJ...910L...8Z,
       author = {{Zhao}, Zhouqiao and {Liu}, Yonggang and {Li}, Weihan and {Liu}, Haobo and {Man}, Kai},
        title = "{Climate Change of over 20 {\textdegree}C Induced by Continental Movement on a Synchronously Rotating Exoplanet}",
      journal = {\apjl},
         year = 2021,
        month = mar,
       volume = {910},
       number = {1},
          eid = {L8},
        pages = {L8},
          doi = {10.3847/2041-8213/abebe6},
       adsurl = {https://ui.adsabs.harvard.edu/abs/2021ApJ...910L...8Z}
}

@ARTICLE{2022GeoRL..4995748O,
       author = {{Olson}, Stephanie and {Jansen}, Malte F. and {Abbot}, Dorian S. and {Halevy}, Itay and {Goldblatt}, Colin},
        title = "{The Effect of Ocean Salinity on Climate and Its Implications for Earth's Habitability}",
      journal = {\grl},
         year = 2022,
        month = may,
       volume = {49},
       number = {10},
          eid = {e95748},
        pages = {e95748},
          doi = {10.1029/2021GL095748},
archivePrefix = {arXiv},
       eprint = {2205.06785},
 primaryClass = {astro-ph.EP},
       adsurl = {https://ui.adsabs.harvard.edu/abs/2022GeoRL..4995748O}
}

@ARTICLE{2019ApJ...884...75D,
       author = {{Del Genio}, Anthony D. and {Kiang}, Nancy Y. and {Way}, Michael J. and {Amundsen}, David S. and {Sohl}, Linda E. and {Fujii}, Yuka and {Chandler}, Mark and {Aleinov}, Igor and {Colose}, Christopher M. and {Guzewich}, Scott D. and {Kelley}, Maxwell},
        title = "{Albedos, Equilibrium Temperatures, and Surface Temperatures of Habitable Planets}",
      journal = {\apj},
         year = 2019,
        month = oct,
       volume = {884},
       number = {1},
          eid = {75},
        pages = {75},
          doi = {10.3847/1538-4357/ab3be8},
archivePrefix = {arXiv},
       eprint = {1812.06606},
 primaryClass = {astro-ph.EP},
       adsurl = {https://ui.adsabs.harvard.edu/abs/2019ApJ...884...75D}
}

@ARTICLE{2020MNRAS.495....1M,
       author = {{Madden}, Jack and {Kaltenegger}, Lisa},
        title = "{How surfaces shape the climate of habitable exoplanets}",
      journal = {\mnras},
         year = 2020,
        month = jun,
       volume = {495},
       number = {1},
        pages = {1-11},
          doi = {10.1093/mnras/staa387},
archivePrefix = {arXiv},
       eprint = {2001.00085},
 primaryClass = {astro-ph.EP},
       adsurl = {https://ui.adsabs.harvard.edu/abs/2020MNRAS.495....1M}
}

@ARTICLE{2020A&A...639A..99E,
       author = {{Eager-Nash}, Jake K. and {Reichelt}, David J. and {Mayne}, Nathan J. and {Hugo Lambert}, F. and {Sergeev}, Denis E. and {Ridgway}, Robert J. and {Manners}, James and {Boutle}, Ian A. and {Lenton}, Timothy M. and {Kohary}, Krisztian},
        title = "{Implications of different stellar spectra for the climate of tidally locked Earth-like exoplanets}",
      journal = {\aap},
         year = 2020,
        month = jul,
       volume = {639},
          eid = {A99},
        pages = {A99},
          doi = {10.1051/0004-6361/202038089},
archivePrefix = {arXiv},
       eprint = {2005.13002},
 primaryClass = {astro-ph.EP},
       adsurl = {https://ui.adsabs.harvard.edu/abs/2020A&A...639A..99E}
}

@ARTICLE{2000Icar..148..508D,
       author = {{Dauphas}, Nicolas and {Robert}, Fran{\c{c}}ois and {Marty}, Bernard},
        title = "{The Late Asteroidal and Cometary Bombardment of Earth as Recorded in Water Deuterium to Protium Ratio}",
      journal = {\icarus},
         year = 2000,
        month = dec,
       volume = {148},
       number = {2},
        pages = {508-512},
          doi = {10.1006/icar.2000.6489},
       adsurl = {https://ui.adsabs.harvard.edu/abs/2000Icar..148..508D}
}

@ARTICLE{2013ApJ...767...54I,
       author = {{Izidoro}, A. and {de Souza Torres}, K. and {Winter}, O.~C. and {Haghighipour}, N.},
        title = "{A Compound Model for the Origin of Earth's Water}",
      journal = {\apj},
         year = 2013,
        month = apr,
       volume = {767},
       number = {1},
          eid = {54},
        pages = {54},
          doi = {10.1088/0004-637X/767/1/54},
archivePrefix = {arXiv},
       eprint = {1302.1233},
 primaryClass = {astro-ph.EP},
       adsurl = {https://ui.adsabs.harvard.edu/abs/2013ApJ...767...54I}
}

@ARTICLE{2015Sci...350..795H,
       author = {{Hallis}, Lydia J. and {Huss}, Gary R. and {Nagashima}, Kazuhide and {Taylor}, G. Jeffrey and {Halld{\'o}rsson}, S{\ae}mundur A. and {Hilton}, David R. and {Mottl}, Michael J. and {Meech}, Karen J.},
        title = "{Evidence for primordial water in Earth{\textquoteright}s deep mantle}",
      journal = {Science},
         year = 2015,
        month = nov,
       volume = {350},
       number = {6262},
        pages = {795-797},
          doi = {10.1126/science.aac4834},
       adsurl = {https://ui.adsabs.harvard.edu/abs/2015Sci...350..795H}
}

@ARTICLE{2008Icar..194...42G,
       author = {{Genda}, Hidenori and {Ikoma}, Masahiro},
        title = "{Origin of the ocean on the Earth: Early evolution of water D/H in a hydrogen-rich atmosphere}",
      journal = {\icarus},
         year = 2008,
        month = mar,
       volume = {194},
       number = {1},
        pages = {42-52},
          doi = {10.1016/j.icarus.2007.09.007},
archivePrefix = {arXiv},
       eprint = {0709.2025},
 primaryClass = {astro-ph},
       adsurl = {https://ui.adsabs.harvard.edu/abs/2008Icar..194...42G}
}

@ARTICLE{2018AsBio..18..469R,
       author = {{Rushby}, Andrew J. and {Johnson}, Martin and {Mills}, Benjamin J.~W. and {Watson}, Andrew J. and {Claire}, Mark W.},
        title = "{Long-Term Planetary Habitability and the Carbonate-Silicate Cycle}",
      journal = {Astrobiology},
         year = 2018,
        month = may,
       volume = {18},
       number = {5},
        pages = {469-480},
          doi = {10.1089/ast.2017.1693},
archivePrefix = {arXiv},
       eprint = {1805.06943},
 primaryClass = {astro-ph.EP},
       adsurl = {https://ui.adsabs.harvard.edu/abs/2018AsBio..18..469R}
}

@article{von2003biogenic,
  title={Biogenic enhancement of weathering and the stability of the ecosphere},
  author={Von Bloh, W and Franck, S and Bounama, C and Schellnhuber, H-J},
  journal={Geomicrobiology Journal},
  volume={20},
  number={5},
  pages={501--511},
  year={2003},
  publisher={Taylor \& Francis}
}

@ARTICLE{2000TellB..52...94F,
       author = {{Franck}, S. and {Block}, A. and {Von Bloh}, W. and {Bounama}, C. and {Schellnhuber}, H.~J. and {Svirezhev}, Y.},
        title = "{Reduction of biosphere life span as a consequence of geodynamics}",
      journal = {Tellus Series B Chemical and Physical Meteorology B},
         year = 2000,
        month = jan,
       volume = {52},
       number = {1},
        pages = {94-107},
          doi = {10.3402/tellusb.v52i1.16085},
       adsurl = {https://ui.adsabs.harvard.edu/abs/2000TellB..52...94F}
}

@ARTICLE{2023IJAsB..22..272M,
       author = {{Mello}, Fernando de Sousa and {Fria{\c{c}}a}, Am{\^a}ncio C{\'e}sar Santos},
        title = "{Planetary geodynamics and age constraints on circumstellar habitable zones around main sequence stars}",
      journal = {International Journal of Astrobiology},
         year = 2023,
        month = aug,
       volume = {22},
       number = {4},
        pages = {272-316},
          doi = {10.1017/S1473550423000083},
       adsurl = {https://ui.adsabs.harvard.edu/abs/2023IJAsB..22..272M}
}

@ARTICLE{1982Natur.296..561L,
       author = {{Lovelock}, J.~E. and {Whitfield}, M.},
        title = "{Life span of the biosphere}",
      journal = {\nat},
         year = 1982,
        month = apr,
       volume = {296},
        pages = {561-563},
          doi = {10.1038/296561a0},
       adsurl = {https://ui.adsabs.harvard.edu/abs/1982Natur.296..561L}
}

@ARTICLE{1992Natur.360..721C,
       author = {{Caldeira}, Ken and {Kasting}, James F.},
        title = "{The life span of the biosphere revisited}",
      journal = {\nat},
         year = 1992,
        month = dec,
       volume = {360},
       number = {6406},
        pages = {721-723},
          doi = {10.1038/360721a0},
       adsurl = {https://ui.adsabs.harvard.edu/abs/1992Natur.360..721C}
}

@ARTICLE{2023MNRAS.524.1491L,
       author = {{Liu}, Binghan and {Marsh}, Daniel R. and {Walsh}, Catherine and {Cooke}, Greg},
        title = "{Higher water loss on Earth-like exoplanets in eccentric orbits}",
      journal = {\mnras},
         year = 2023,
        month = sep,
       volume = {524},
       number = {1},
        pages = {1491-1502},
          doi = {10.1093/mnras/stad1828},
archivePrefix = {arXiv},
       eprint = {2306.10958},
 primaryClass = {astro-ph.EP},
       adsurl = {https://ui.adsabs.harvard.edu/abs/2023MNRAS.524.1491L}
}

@ARTICLE{2011ApJ...729...54C,
       author = {{Cowan}, Nicolas B. and {Agol}, Eric},
        title = "{The Statistics of Albedo and Heat Recirculation on Hot Exoplanets}",
      journal = {\apj},
         year = 2011,
        month = mar,
       volume = {729},
       number = {1},
          eid = {54},
        pages = {54},
          doi = {10.1088/0004-637X/729/1/54},
archivePrefix = {arXiv},
       eprint = {1001.0012},
 primaryClass = {astro-ph.EP},
       adsurl = {https://ui.adsabs.harvard.edu/abs/2011ApJ...729...54C}
}

@ARTICLE{2014RSPTA.37230084F,
       author = {{Forget}, F. and {Leconte}, J.},
        title = "{Possible climates on terrestrial exoplanets}",
      journal = {Philosophical Transactions of the Royal Society of London Series A},
         year = 2014,
        month = mar,
       volume = {372},
       number = {2014},
        pages = {20130084-20130084},
          doi = {10.1098/rsta.2013.0084},
archivePrefix = {arXiv},
       eprint = {1311.3101},
 primaryClass = {astro-ph.EP},
       adsurl = {https://ui.adsabs.harvard.edu/abs/2014RSPTA.37230084F}
}

@ARTICLE{2022MNRAS.513.2761M,
       author = {{Macdonald}, Evelyn and {Paradise}, Adiv and {Menou}, Kristen and {Lee}, Christopher},
        title = "{Climate uncertainties caused by unknown land distribution on habitable M-Earths}",
      journal = {\mnras},
         year = 2022,
        month = jun,
       volume = {513},
       number = {2},
        pages = {2761-2769},
          doi = {10.1093/mnras/stac1040},
archivePrefix = {arXiv},
       eprint = {2110.04310},
 primaryClass = {astro-ph.EP},
       adsurl = {https://ui.adsabs.harvard.edu/abs/2022MNRAS.513.2761M}
}

@ARTICLE{2015ApJ...813L...3K,
       author = {{Kasting}, James F. and {Chen}, Howard and {Kopparapu}, Ravi K.},
        title = "{Stratospheric Temperatures and Water Loss from Moist Greenhouse Atmospheres of Earth-like Planets}",
      journal = {\apjl},
         year = 2015,
        month = nov,
       volume = {813},
       number = {1},
          eid = {L3},
        pages = {L3},
          doi = {10.1088/2041-8205/813/1/L3},
archivePrefix = {arXiv},
       eprint = {1510.03527},
 primaryClass = {astro-ph.EP},
       adsurl = {https://ui.adsabs.harvard.edu/abs/2015ApJ...813L...3K}
}

@ARTICLE{2023MNRAS.523..286S,
       author = {{Schulik}, Matth{\"a}us and {Booth}, Richard A.},
        title = "{AIOLOS - A multipurpose 1D hydrodynamics code for planetary atmospheres}",
      journal = {\mnras},
         year = 2023,
        month = jul,
       volume = {523},
       number = {1},
        pages = {286-304},
          doi = {10.1093/mnras/stad1251},
archivePrefix = {arXiv},
       eprint = {2207.07144},
 primaryClass = {astro-ph.EP},
       adsurl = {https://ui.adsabs.harvard.edu/abs/2023MNRAS.523..286S}
}

@ARTICLE{2018PNAS..115..260D,
       author = {{Dong}, Chuanfei and {Jin}, Meng and {Lingam}, Manasvi and {Airapetian}, Vladimir S. and {Ma}, Yingjuan and {van der Holst}, Bart},
        title = "{Atmospheric escape from the TRAPPIST-1 planets and implications for habitability}",
      journal = {Proceedings of the National Academy of Science},
         year = 2018,
        month = jan,
       volume = {115},
       number = {2},
        pages = {260-265},
          doi = {10.1073/pnas.1708010115},
archivePrefix = {arXiv},
       eprint = {1705.05535},
 primaryClass = {astro-ph.EP},
       adsurl = {https://ui.adsabs.harvard.edu/abs/2018PNAS..115..260D}
}

@ARTICLE{2021MNRAS.500.2197Z,
       author = {{Zilinskas}, Mantas and {Miguel}, Yamila and {Lyu}, Yipeng and {Bax}, Morris},
        title = "{Temperature inversions on hot super-Earths: the case of CN in nitrogen-rich atmospheres}",
      journal = {\mnras},
         year = 2021,
        month = jan,
       volume = {500},
       number = {2},
        pages = {2197-2208},
          doi = {10.1093/mnras/staa3415},
archivePrefix = {arXiv},
       eprint = {2010.15152},
 primaryClass = {astro-ph.EP},
       adsurl = {https://ui.adsabs.harvard.edu/abs/2021MNRAS.500.2197Z}
}

@ARTICLE{2019MinDe..54..485L,
       author = {{Large}, Ross R. and {Mukherjee}, Indrani and {Gregory}, Dan and {Steadman}, Jeff and {Corkrey}, Ross and {Danyushevsky}, Leonid V.},
        title = "{Atmosphere oxygen cycling through the Proterozoic and Phanerozoic}",
      journal = {Mineralium Deposita},
         year = 2019,
        month = apr,
       volume = {54},
       number = {4},
        pages = {485-506},
          doi = {10.1007/s00126-019-00873-9},
       adsurl = {https://ui.adsabs.harvard.edu/abs/2019MinDe..54..485L}
}

@ARTICLE{2009RvGeo..47.1004F,
       author = {{Fueglistaler}, S. and {Dessler}, A.~E. and {Dunkerton}, T.~J. and {Folkins}, I. and {Fu}, Q. and {Mote}, P.~W.},
        title = "{Tropical tropopause layer}",
      journal = {Reviews of Geophysics},
         year = 2009,
        month = mar,
       volume = {47},
       number = {1},
          eid = {RG1004},
        pages = {RG1004},
          doi = {10.1029/2008RG000267},
       adsurl = {https://ui.adsabs.harvard.edu/abs/2009RvGeo..47.1004F}
}

@ARTICLE{2020AsBio..20..628P,
       author = {{Planavsky}, Noah J. and {Reinhard}, Christopher T. and {Isson}, Terry T. and {Ozaki}, Kazumi and {Crockford}, Peter W.},
        title = "{Large Mass-Independent Oxygen Isotope Fractionations in Mid-Proterozoic Sediments: Evidence for a Low-Oxygen Atmosphere?}",
      journal = {Astrobiology},
         year = 2020,
        month = may,
       volume = {20},
       number = {5},
        pages = {628-636},
          doi = {10.1089/ast.2019.2060},
       adsurl = {https://ui.adsabs.harvard.edu/abs/2020AsBio..20..628P}
}

@ARTICLE{2017A&A...601A.120B,
       author = {{Boutle}, Ian A. and {Mayne}, Nathan J. and {Drummond}, Benjamin and {Manners}, James and {Goyal}, Jayesh and {Hugo Lambert}, F. and {Acreman}, David M. and {Earnshaw}, Paul D.},
        title = "{Exploring the climate of Proxima B with the Met Office Unified Model}",
      journal = {\aap},
         year = 2017,
        month = may,
       volume = {601},
          eid = {A120},
        pages = {A120},
          doi = {10.1051/0004-6361/201630020},
archivePrefix = {arXiv},
       eprint = {1702.08463},
 primaryClass = {astro-ph.EP},
       adsurl = {https://ui.adsabs.harvard.edu/abs/2017A&A...601A.120B}
}

@ARTICLE{2017PNAS..114.1811G,
       author = {{Gumsley}, Ashley P. and {Chamberlain}, Kevin R. and {Bleeker}, Wouter and {S{\"o}derlund}, Ulf and {de Kock}, Michiel O. and {Larsson}, Emilie R. and {Bekker}, Andrey},
        title = "{Timing and tempo of the Great Oxidation Event}",
      journal = {Proceedings of the National Academy of Science},
         year = 2017,
        month = feb,
       volume = {114},
       number = {8},
        pages = {1811-1816},
          doi = {10.1073/pnas.1608824114},
       adsurl = {https://ui.adsabs.harvard.edu/abs/2017PNAS..114.1811G}
}

@ARTICLE{2023RSOS...1030056J,
       author = {{Ji}, A. and {Kasting}, J.~F. and {Cooke}, G.~J. and {Marsh}, D.~R. and {Tsigaridis}, K.},
        title = "{Comparison between ozone column depths and methane lifetimes computed by one- and three-dimensional models at different atmospheric O$_{2}$ levels}",
      journal = {Royal Society Open Science},
         year = 2023,
        month = may,
       volume = {10},
       number = {5},
          eid = {230056},
        pages = {230056},
          doi = {10.1098/rsos.230056},
       adsurl = {https://ui.adsabs.harvard.edu/abs/2023RSOS...1030056J}
}

@ARTICLE{2017ApJS..231...12W,
       author = {{Way}, M.~J. and {Aleinov}, I. and {Amundsen}, David S. and {Chandler}, M.~A. and {Clune}, T.~L. and {Del Genio}, A.~D. and {Fujii}, Y. and {Kelley}, M. and {Kiang}, N.~Y. and {Sohl}, L. and {Tsigaridis}, K.},
        title = "{Resolving Orbital and Climate Keys of Earth and Extraterrestrial Environments with Dynamics (ROCKE-3D) 1.0: A General Circulation Model for Simulating the Climates of Rocky Planets}",
      journal = {\apjs},
         year = 2017,
        month = jul,
       volume = {231},
       number = {1},
          eid = {12},
        pages = {12},
          doi = {10.3847/1538-4365/aa7a06},
archivePrefix = {arXiv},
       eprint = {1701.02360},
 primaryClass = {astro-ph.EP},
       adsurl = {https://ui.adsabs.harvard.edu/abs/2017ApJS..231...12W}
}

@ARTICLE{2018ApJ...854..171L,
       author = {{Lewis}, Neil T. and {Lambert}, F. Hugo and {Boutle}, Ian A. and {Mayne}, Nathan J. and {Manners}, James and {Acreman}, David M.},
        title = "{The Influence of a Substellar Continent on the Climate of a Tidally Locked Exoplanet}",
      journal = {\apj},
         year = 2018,
        month = feb,
       volume = {854},
       number = {2},
          eid = {171},
        pages = {171},
          doi = {10.3847/1538-4357/aaad0a},
archivePrefix = {arXiv},
       eprint = {1802.00378},
 primaryClass = {astro-ph.EP},
       adsurl = {https://ui.adsabs.harvard.edu/abs/2018ApJ...854..171L}
}

@ARTICLE{2022CliPa..18.2421Y,
       author = {{Yassin Jaziri}, Adam and {Charnay}, Benjamin and {Selsis}, Franck and {Leconte}, J{\'e}r{\'e}my and {Lef{\`e}vre}, Franck},
        title = "{Dynamics of the Great Oxidation Event from a 3D photochemical-climate model}",
      journal = {Climate of the Past},
         year = 2022,
        month = oct,
       volume = {18},
       number = {10},
        pages = {2421-2447},
          doi = {10.5194/cp-18-2421-2022},
archivePrefix = {arXiv},
       eprint = {2212.01389},
 primaryClass = {astro-ph.EP},
       adsurl = {https://ui.adsabs.harvard.edu/abs/2022CliPa..18.2421Y}
}

@ARTICLE{2019JGRD..12412380G,
       author = {{Gettelman}, A. and {Mills}, M.~J. and {Kinnison}, D.~E. and {Garcia}, R.~R. and {Smith}, A.~K. and {Marsh}, D.~R. and {Tilmes}, S. and {Vitt}, F. and {Bardeen}, C.~G. and {McInerny}, J. and {Liu}, H. -L. and {Solomon}, S.~C. and {Polvani}, L.~M. and {Emmons}, L.~K. and {Lamarque}, J. -F. and {Richter}, J.~H. and {Glanville}, A.~S. and {Bacmeister}, J.~T. and {Phillips}, A.~S. and {Neale}, R.~B. and {Simpson}, I.~R. and {DuVivier}, A.~K. and {Hodzic}, A. and {Randel}, W.~J.},
        title = "{The Whole Atmosphere Community Climate Model Version 6 (WACCM6)}",
      journal = {Journal of Geophysical Research (Atmospheres)},
         year = 2019,
        month = dec,
       volume = {124},
       number = {23},
        pages = {12,380-12,403},
          doi = {10.1029/2019JD030943},
       adsurl = {https://ui.adsabs.harvard.edu/abs/2019JGRD..12412380G}
}

@ARTICLE{2022MNRAS.517.2383B,
       author = {{Braam}, Marrick and {Palmer}, Paul I. and {Decin}, Leen and {Ridgway}, Robert J. and {Zamyatina}, Maria and {Mayne}, Nathan J. and {Sergeev}, Denis E. and {Abraham}, N. Luke},
        title = "{Lightning-induced chemistry on tidally-locked Earth-like exoplanets}",
      journal = {\mnras},
         year = 2022,
        month = dec,
       volume = {517},
       number = {2},
        pages = {2383-2402},
          doi = {10.1093/mnras/stac2722},
archivePrefix = {arXiv},
       eprint = {2209.12502},
 primaryClass = {astro-ph.EP},
       adsurl = {https://ui.adsabs.harvard.edu/abs/2022MNRAS.517.2383B}
}

@article{ji2023comparison,
  title={Comparison between ozone column depths and methane lifetimes computed by one-and three-dimensional models at different atmospheric O2 levels},
  author={Ji, A and Kasting, JF and Cooke, GJ and Marsh, DR and Tsigaridis, K},
  journal={Royal Society Open Science},
  volume={10},
  number={5},
  pages={230056},
  year={2023},
  publisher={The Royal Society}
}

@ARTICLE{1949QJRMS..75..351B,
       author = {{Brewer}, A.~W.},
        title = "{Evidence for a world circulation provided by the measurements of helium and water vapour distribution in the stratosphere}",
      journal = {Quarterly Journal of the Royal Meteorological Society},
         year = 1949,
        month = oct,
       volume = {75},
       number = {326},
        pages = {351-363},
          doi = {10.1002/qj.49707532603},
       adsurl = {https://ui.adsabs.harvard.edu/abs/1949QJRMS..75..351B}
}

@ARTICLE{1926RSPSA.110..660D,
       author = {{Dobson}, G.~M.~B. and {Harrison}, D.~N.},
        title = "{Measurements of the Amount of Ozone in the Earth's Atmosphere and Its Relation to Other Geophysical Conditions}",
      journal = {Proceedings of the Royal Society of London Series A},
         year = 1926,
        month = apr,
       volume = {110},
       number = {756},
        pages = {660-693},
          doi = {10.1098/rspa.1926.0040},
       adsurl = {https://ui.adsabs.harvard.edu/abs/1926RSPSA.110..660D}
}

@ARTICLE{2014RvGeo..52..157B,
       author = {{Butchart}, Neal},
        title = "{The Brewer-Dobson circulation}",
      journal = {Reviews of Geophysics},
         year = 2014,
        month = jun,
       volume = {52},
       number = {2},
        pages = {157-184},
          doi = {10.1002/2013RG000448},
       adsurl = {https://ui.adsabs.harvard.edu/abs/2014RvGeo..52..157B}
}

\end{document}